\documentclass[12pt]{iopart}

\usepackage{hyperref}
\usepackage{graphicx}
\usepackage{listings}
\usepackage{bm}
\usepackage[pagewise]{lineno}
\begin{document}

\title[Identifying correlations in LIGO using lasso]{Identifying correlations between LIGO's astronomical range and auxiliary sensors using lasso regression}

% addresses now follow section 3.2. Authors’ names and addresses in IOP guidelines PDF

\newcommand{\GWPAC}{\address{$^1$ Gravitational Wave Physics and Astronomy Center, California State University Fullerton, Fullerton, CA 92834, USA}}

\author{Marissa Walker$^1$, Alfonso F. Agnew$^{1,2}$, Jeffrey Bidler$^1$, Andrew~Lundgren$^3$, Alexandra Macedo$^1$, Duncan Macleod$^4$, T.J. Massinger$^5$, Oliver Patane$^1$ and Joshua R. Smith$^1$}

\GWPAC
\address{$^2$ Department of Mathematics, California State University Fullerton, Fullerton, CA 92834, USA}
\address{$^3$ Institute of Cosmology and Gravitation, University of Portsmouth, Dennis Sciama Building, Burnaby Road, Portsmouth, PO1 3FX, United Kingdom}
\address{$^4$ School of Physics and Astronomy, Cardiff University, Cardiff CF24 3AA, United Kingdom}
\address{$^5$ LIGO, California Institute of Technology, Pasadena, CA 91125, USA}

\ead{mwalker@fullerton.edu}

\begin{abstract}
The range to which the Laser Interferometer Gravitational-Wave Observatory (LIGO) can observe astrophysical systems varies over time, limited by noise in the instruments and their environments. Identifying and removing the sources of noise that limit LIGO's range enables higher signal-to-noise observations and increases the number of observations. The LIGO observatories are continuously monitored by hundreds of thousands of auxiliary channels that may contain information about these noise sources. This paper describes an algorithm that uses linear regression, namely  lasso (least absolute shrinkage and selection operator) regression, to analyze all of these channels and identify a small subset of them that can be used to reconstruct variations in LIGO's astrophysical range. Exemplary results of the application of this method to three different periods of LIGO Livingston data are presented, along with computational performance and current limitations. 
\end{abstract}

\submitto{\CQG}
\maketitle

\section{Introduction}

The Laser Interferometer Gravitational-Wave Observatory (LIGO)~\cite{aligo} dramatically opened the field of gravitational-wave astronomy in 2015 by observing the merging binary black hole system GW150914~\cite{gw150914}. LIGO and Virgo~\cite{virgo2014} have since observed several more black hole mergers~\cite{o1bbh,gw151226,gw170104,gw170608,gw170814}. In 2017, LIGO and Virgo debuted gravitational-wave multi-messenger astronomy with the discovery of the binary neutron star merger GW170817~\cite{gw170817} and rapidly reported its source's location, leading to electromagnetic observation by the astronomical community~\cite{gw170817mma}. These breakthroughs, measurements of the incredibly minute changes in spacetime caused by astrophysical gravitational waves, were the direct result of decades of research and development leading to observatories with unprecedented sensitivity. 

LIGO Hanford (H1) and LIGO Livingston (L1) are highly complex kilometer-scale laser interferometric detectors~\cite{aligo}. Their sensitivity to gravitational waves is limited by sources of noise in the instruments and their environments. A common benchmark of this sensitivity is the ``binary-neutron-star (BNS) range", which is defined as the distance at which a single detector could observe the coalescence of a pair of 1.4 solar mass neutron stars with a signal-to-noise ratio of 8~\cite{Finn1993, observingscenarios}. During LIGO's first two observing runs, O1 and O2, the range of each detector varied significantly on minute and hourly timescales, often by tens of percent. Identifying and removing the causes of this variation to maximize the range would lead to higher signal-to-noise observations and would dramatically increase the number of observations because the volume of space accessible is proportional to the cube of the range. 

Operating the LIGO detectors at the sensitivity needed for gravitational-wave detection requires a large number of control loops and sensors to precisely control and monitor the instrumentation and environment. The channels that continually record information from these sensors enable investigations into problematic variations in the sensitivity of the detector. However, due to the complexity of the system, the number of auxiliary sensor channels is roughly 500,000. Searching through such a large amount of data to determine which sensors are most highly correlated with variations in the range is a formidable yet important challenge.

Previous work to address this challenge includes the Non-stationary Noise Analysis (NonNA)~\cite{VajenteNonNA,vajente2008analysis}, based on ordinary least squares analysis performed in multiple rounds to select auxiliary channels with large contributions to a primary time series, such as the BNS range. In brief, that method works by performing linear regression of the primary time series with a number of auxiliary channels, and the channel whose fit has the smallest residual is noted and its contribution subtracted away from the main time series. This process is iterated, resulting in a ranked list of auxiliary channels that are strongly correlated with the main time series.

This paper describes an algorithm that uses regularized linear regression to analyze hundreds of thousands of channels over long time periods (typically 24 hours) and identify a small subset of those channels that can be used to reconstruct the variations in the BNS range. Further, it uses correlation analysis to identify groups of channels with similar behavior to those chosen for the model. This results in a relatively short list of channels that can be investigated further to help identify the root causes of variations in the detector's range.

\section{Linear Regression Methods}\label{sec:methods}

Regression methods can be employed to discern relationships between the auxiliary channels and the range, by treating the channel data as input variables to model the range as the output. Each channel provides a time series vector $x_i$ as input, and so we let $X = \left( X_{ij} \right)$ denote the matrix whose $j$-th column denotes the time series for channel $j$. We then seek a linear regression model of the form 
\begin{equation}
f(X) = \sum_{j}X_{ij}\beta_j.
\end{equation}
Here, $\bm\beta = \left(\beta_j\right)$ is the regression coefficient vector, to be determined by the regression scheme,  where the $j$-th component is the regression coefficient for channel $j.$ Ordinary least squares (OLS) regression corresponds to choosing $\bm\beta$ to minimize the error in the output of the model, measured by the residual sum of squares~\cite{hastie_09_elements-of.statistical-learning},
\begin{equation}\label{equation:rss}
RSS(\bm\beta) = ||\mathbf{y} - \mathbf{X}\bm{\beta}||^2 = \sum_{i}||y_i - \sum_{j}X_{ij}\beta_j||^2,
\end{equation}
where $\mathbf{y} = \left( y_i \right)$ represents the time series for the range.

\begin{figure*}[ht]
\begin{center}
\includegraphics[width=1.0\linewidth]{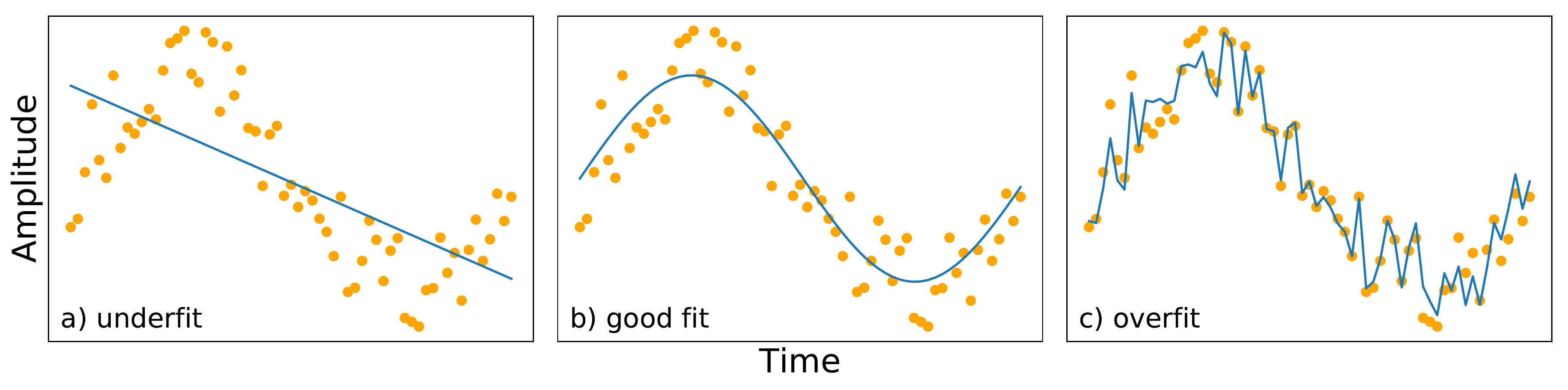}
\caption{An example of different levels of fitting, underfit on the left using first order ordinary least squares, well-fit in the center, and overfit on the right. \label{fig:fit-level}} 
\end{center}
\end{figure*}

A large variety of regression methods exist today, and the selection of an appropriate method depends on the application. One challenge of choosing a regression technique is finding the right balance between two competing virtues: variance and bias. On one hand, a desirable model accurately fits the variations in the data, but on the other hand, the model is more useful if it is simple and interpretable. Placing too much emphasis on the variance can lead to \emph{overfitting} (see Figure~\ref{fig:fit-level}), in which the model fits random statistical variations in the output data, rather than the targeted variations. For large data sets in particular, this may result in a highly complex model that may be difficult to interpret, or even misleading. To simplify a model, the amount of variation can be reduced by introducing a bias toward certain input variables or by eliminating a subset of the inputs altogether.  

In our case, because LIGO is such a complex instrument, the number of auxiliary channels is so large, and scientists' time to fix problems is so limited, interpretability of the model is of primary importance. The goal is to identify the channels that correlate significantly with the range and eliminate the channels that do not. We expect the number of important channels to be a very small percentage of the total --- more simply put, we are searching for the proverbial ``needle in the haystack''. 

The process of determining which input variables to keep in a model is called \emph{feature selection}. One approach to feature selection is regularization of the allowed values of coefficients by introducing a penalty term $P_\alpha$ to the minimization function used in the regression analysis, in addition to the residual sum of squares (Equation \ref{equation:rss}):

\begin{equation}\label{equation:rsswithpenalty}
||\mathbf{y} - \mathbf{X}\bm{\beta}||^2 + P_{\alpha}(\bm\beta),
\end{equation}

where $\alpha$ is a tunable parameter that penalizes overfitting by reducing or eliminating the contributions from unnecessary input variables. Two common and powerful approaches to regularized regression are ridge regression~\cite{ridge} and lasso regression~\cite{tibshirani1996regression}, which correspond to penalty terms $P^R_{\alpha}(\bm\beta )$ and $P^L_{\alpha}(\bm\beta )$ respectively:

\begin{equation}\label{equation:penalty}P^R_{\alpha}(\bm\beta ) = \alpha \sum_{i = 0}^p \beta_i^2 ,\;\;\;\;
P^L_{\alpha}(\bm\beta ) = \alpha \sum_{i = 0}^p |\beta_i|
\end{equation}

While ridge regression merely reduces the values of the coefficients of the less desirable explanatory variables, lasso regression tends to shrink coefficients to zero, thereby completely eliminating those variables from the model. The difference in behavior between these two regression methods can be understood geometrically in the vector space defined by the possible values of the coefficients. The regularization terms constrain the coefficient vector within a neighborhood whose shape varies depending on the choice of geometry of the vector space, which is determined by the definition of the norm (or length) of a vector. Ridge regression corresponds to the $L^2$ norm while lasso regression is defined using the $L^1$ norm. In familiar Euclidean geometry, which uses the $L^2$ norm, the length of a vector $x$ in two dimensions is $\sqrt[]{x_1^2 + x_2^2}$. In this geometry, vectors with equal length trace out a circle around the origin. The $L^1$ norm is the sum of the absolute values of the components of the vector. Vectors of equal length in a two-dimensional space in this geometry create a diamond-shaped region around the origin, with corners on the axes. 

\begin{figure*}[ht]
\begin{center}
\includegraphics[width=0.9\linewidth]{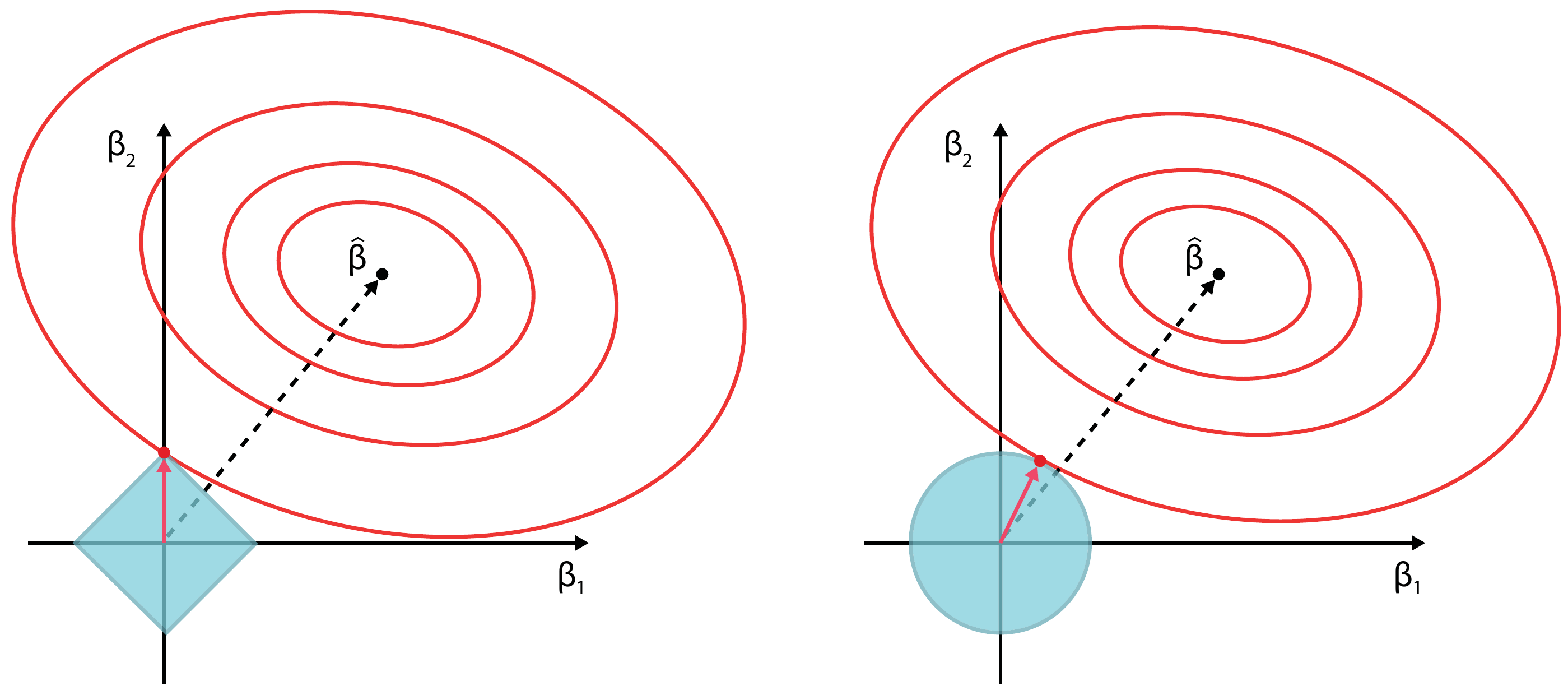}
\caption{A geometric illustration of how lasso sets certain auxiliary channel coefficients to zero (left) compared with ridge regression (right), illustrated in a two-dimensional coefficient space. The dashed arrow represents the coefficient vector that would minimize the least squares error, and the ellipses surrounding this point are contours of equal error. The blue diamond and circle are the constraints placed on the model by lasso and ridge regression, respectively. By forcing the coefficient vector to lie within this region, the final coefficients chosen are the red arrows. While the final coefficients are smaller than ordinary least squares in both cases, the choice of coefficients in lasso sets $\beta_1$ equal to zero, eliminating it from the model. \label{fig:L1vL2}} 
\end{center}
\end{figure*}

Figure~\ref{fig:L1vL2} illustrates the two-dimensional case of the $L^1$ and $L^2$ neighborhoods created by the penalty terms in Equation \ref{equation:penalty}, along with contour ellipses representing constant least squares error values.
Minimizing the cost function in Equation~\ref{equation:rsswithpenalty} is geometrically equivalent to finding the lowest error ellipse that intersects the neighborhood defined by the penalty. The point of this intersection determines the regression coefficients. 
Intersections are likely to occur at the corners in the $L^1$ case, and since the corners lie on axes, the result sets one component of $\bm\beta$ to zero. This feature of lasso regression makes it an ideal tool for our case, as it can eliminate many of the auxiliary channels to create a simple model based on the most important ones.

\section{Algorithm}

We first developed a simple algorithm to identify correlations between the range of each LIGO detector and thousands of LIGO's auxiliary channels, based on Spearman and Pearson correlation coefficients. The inputs to this algorithm are the binary neutron star range of LIGO, sampled once per minute, and minute trends of all or a subset of LIGO's auxiliary channels. It calculates the Spearman and Pearson correlation coefficient between each auxiliary channel and the range, and then produces a webpage that ranks the channels by their highest overall correlation coefficient. For each channel, it also provides graphs such as time series overlays and correlation scatter plots. This algorithm, though very simple, is capable of identifying groups of channels that are related to the sensitivity fluctuations of LIGO. 

However, this approach does not often lead to easily interpretable results. With so many auxiliary channels, results with high correlation coefficient are often those witnessing the same phenomena. Thus, only the set of channels that witness the leading driver of the range fluctuations are highlighted, and those that may relate to subtler variations in the range are buried in a long list due to their lower correlation coefficients. 

For these reasons, we chose to use lasso regression to create a model for the range using a small ranked list of auxiliary channels, rather than an exhaustive list~\cite{tibshirani1996regression}. As we describe below, lasso regression is able to identify the channels related to the noise sources driving range variations, and can be tuned to use only about 10 channels to construct an accurate model of the range, even when provided with thousands of input auxiliary channels. Furthermore, the Pearson correlation coefficient can be used to compare each of the small set of channels selected by lasso with all other channels in order to identify ``clusters" of channels that are highly correlated with the channels used in the model.

\begin{figure*}[ht]
\begin{center}
\includegraphics[width=0.9\linewidth]{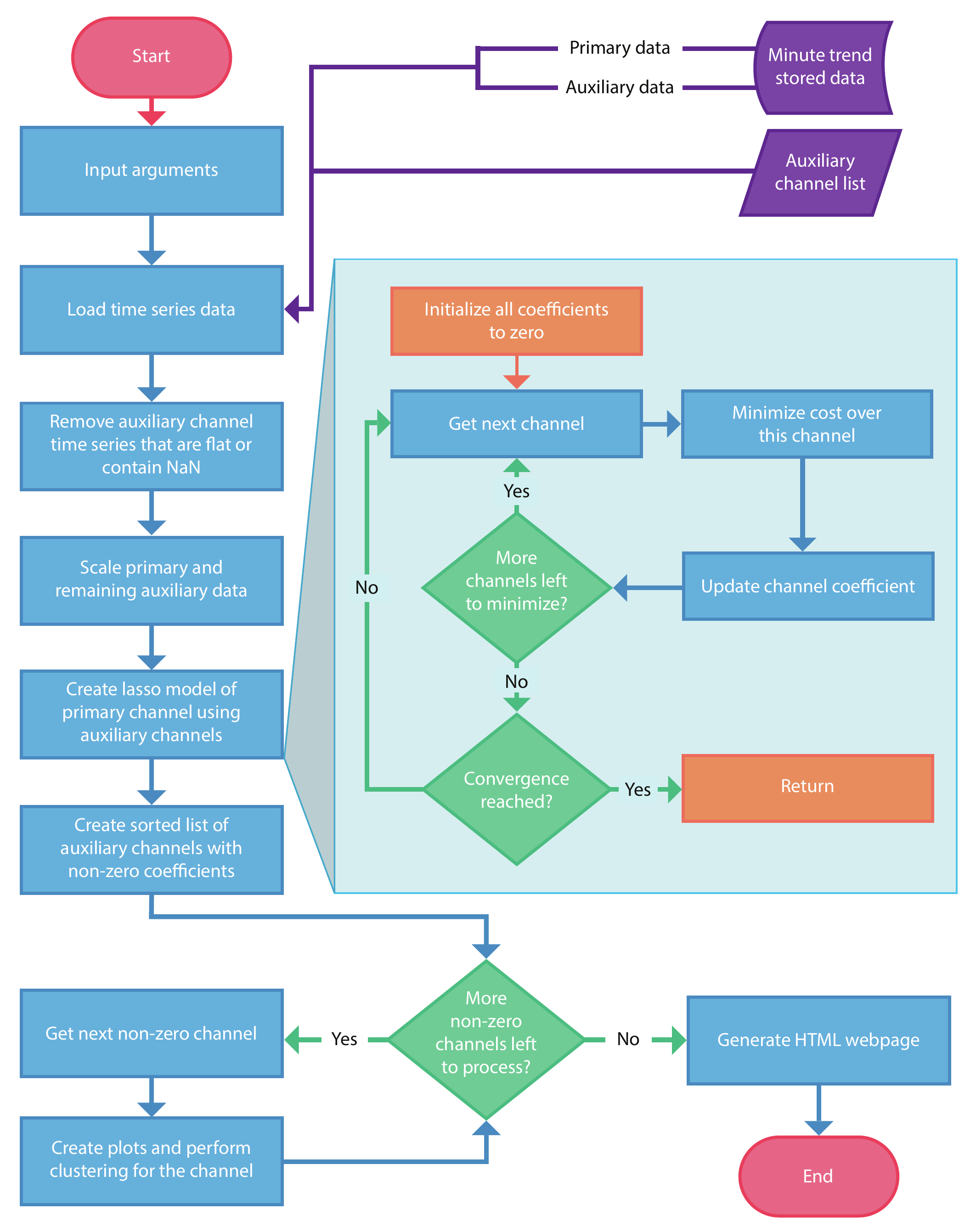}
\caption{Flowchart of the primary operations of the algorithm. Inset: Creation of the lasso model using the scikit-learn lasso method.
 \label{fig:flowchart}} 
\end{center}
\end{figure*}

A flowchart of the algorithm is shown in Figure~\ref{fig:flowchart}. The algorithm is implemented in software using Python and relies on GWpy~\cite{gwpy}, a package for gravitational-wave astrophysics, and the machine learning package scikit-learn~\cite{scikit-learn}, which contains the functions used to scale the data and create the lasso model.

\subsection{Initialization, data retrieval, and preprocessing}
The software is initialized from the command-line or as part of a workflow with input arguments provided by the user (see Table \ref{tab:table1}) and a list of the names of auxiliary channels to consider. See the appendix for brief descriptions of the channels mentioned in this paper.

\begin{table}[h!]
  \begin{center}
    %\captionsetup[table]{justification=centering}
    %\captionsetup[table]{position=bottom} %can't get caption to go beneath table
    \caption{Primary input arguments for the algorithm.}
    \label{tab:table1}
    \footnotesize
    \begin{tabular}{l|l|l}
      \textbf{Argument} & \textbf{Description (default)}\\
      %$Argument$ & $Default$ & $Optional$ \\
      \hline
      Interferometer & Which interferometer, H1 or L1, to run on \\
     % Range Channel & DMT-SNSH\_EFFECTIVE\_RANGE\_MPC.mean \\
      Analysis time &  Start and end GPS time of analysis \\
      %End time &  Start GPS time of analysis \\
      Lasso alpha & Alpha parameter for lasso model (0.1) \\
      %Clustering & Yes/No, whether to search for additional correlated channels (Yes) \\
      Cluster threshold & Threshold Pearson coefficient for clustering (.85) \\
%      Number of processors & Number of processors to run on (1) \\
      %Outlier removal & Yes/No, whether to remove outliers (No)\\
      Outlier removal sigma & The number of standard deviations used for outlier removal (3.0)\\ 
    \end{tabular}
  \end{center}
\end{table}

After the input arguments have been processed, the time series for the range channel is retrieved for the time and interferometer requested. LIGO's range data often has large downward excursions from the general slowly-varying trends, corresponding to loud noise transients that affect the range calculation for a short time. Since the causes of these spikes are better analyzed using transient noise analyses than this tool, which targets slower variations in the data, outliers beyond a user-specified number of standard deviations are optionally removed from the range data.

Based on the provided channel list, a time series for each auxiliary channel is also retrieved. Since many of these channels provide a record of instrumental settings that do not change during observing periods, channels that have constant values during the time interval of the analysis are identified (by checking whether the minimum and maximum values in their time series are equal)
and removed. Additionally, channels containing any NaN values (Not a Number, used to indicate zeros, infinities or other non-numerical values) are removed. With these channels removed, both the range time series and the auxiliary channel time series are scaled to have zero mean and unit variance using scikit-learn's preprocessing module. 

\subsection{Lasso model}

\begin{figure}[ht] 
  \label{stuff} 
  \begin{minipage}{0.8\linewidth}
    \centering
    \includegraphics[width=1.0\linewidth]{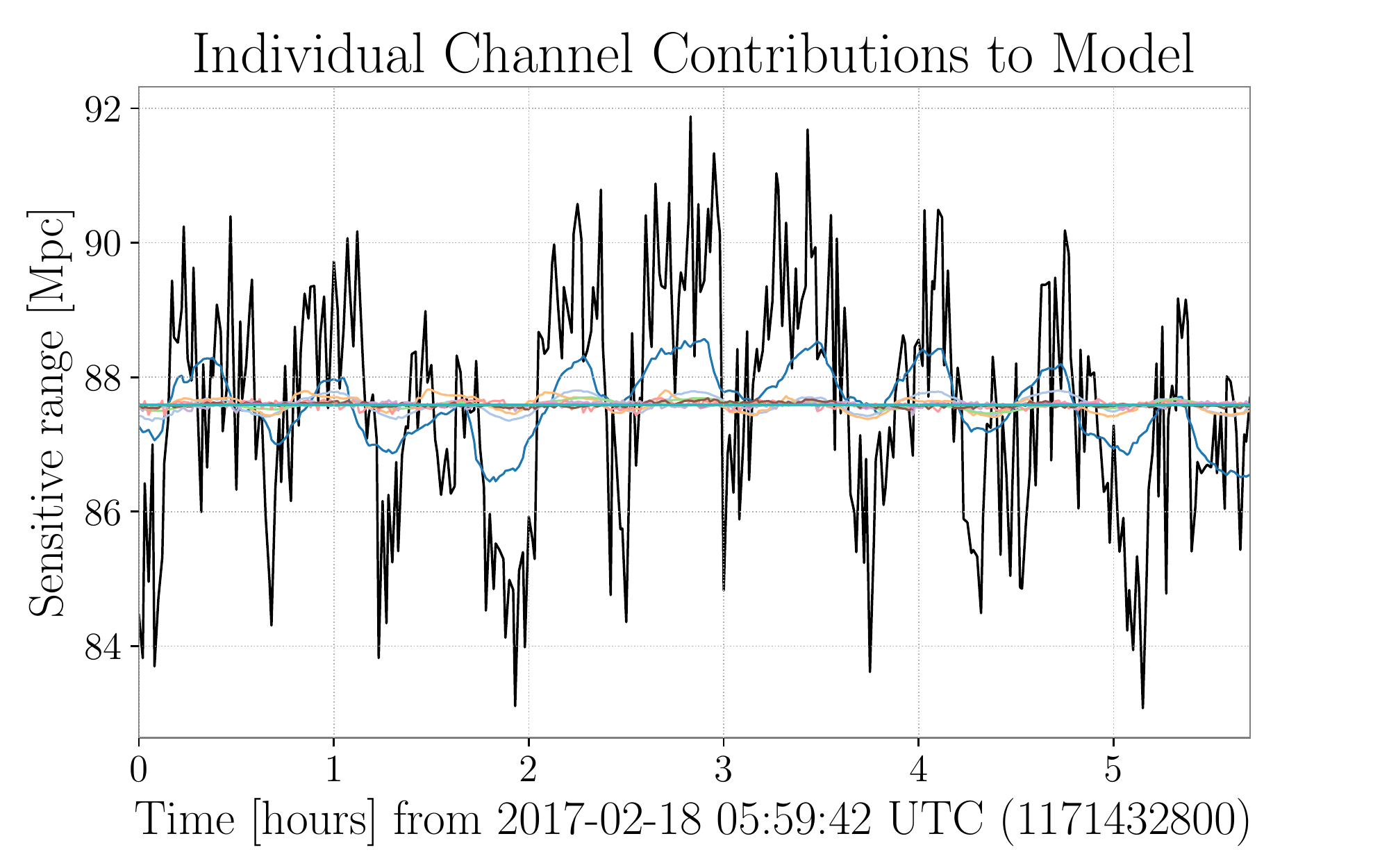}  
    %\vspace{4ex}
  \end{minipage}%%
  \begin{minipage}{0.2\linewidth}
    \centering
    \includegraphics[width=1.0\linewidth]{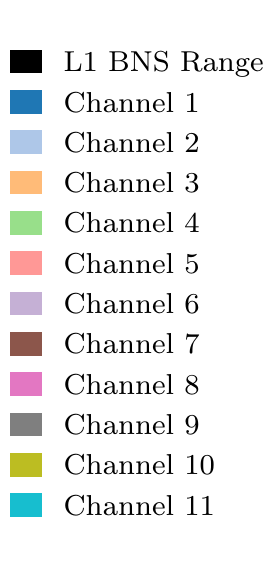} 
    %\vspace{4ex}
  \end{minipage} 
  \begin{minipage}{0.8\linewidth}
    \centering
    \includegraphics[width=1.0\linewidth]{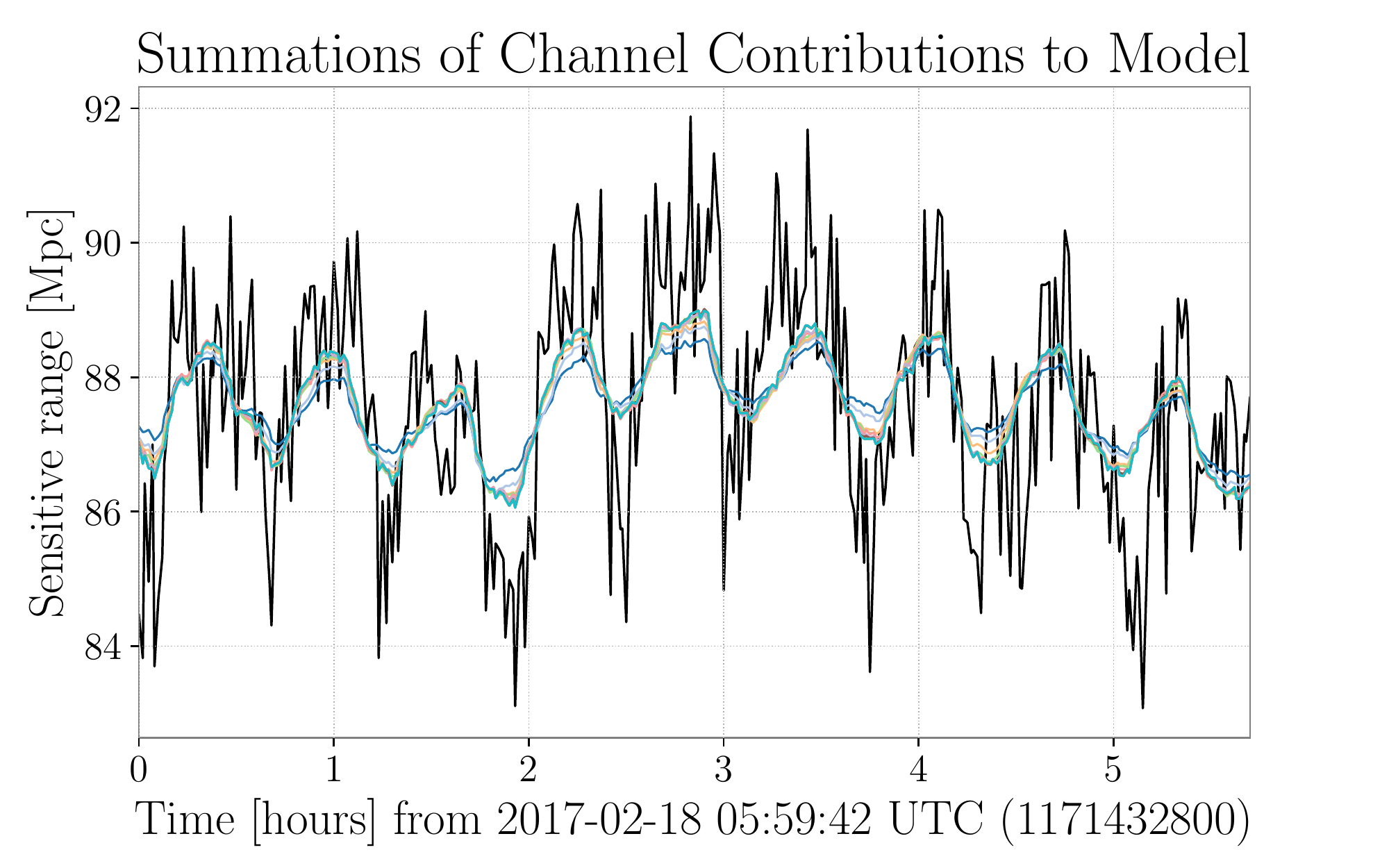} 
    %\vspace{4ex}
  \end{minipage}%% 
  \begin{minipage}{0.2\linewidth}
    \centering
    \includegraphics[width=1.0\linewidth]{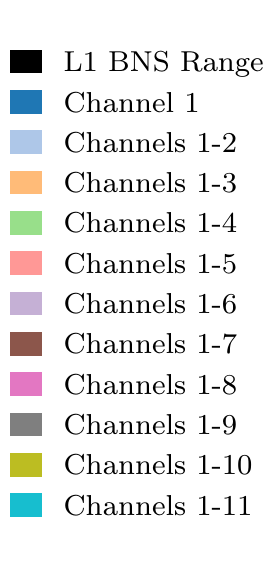} 
    %\vspace{4ex}
  \end{minipage}
  \caption{LIGO range and lasso model comparisons. Top row: individual contributions of auxiliary channels to the lasso model. Bottom row: Cumulative sum contributions of auxiliary channels to the lasso model, starting with the channel with the highest coefficient. \label{fig:modelplots}}
\end{figure}

Using scikit-learn's lasso method, lasso regression is performed to construct a model of the range data using the auxiliary data channels. As described in Section \ref{sec:methods}, the regularization parameter, alpha, determines how many of the channel coefficients will be driven to zero. A larger alpha value will result in fewer channels being used in the model. 
With an appropriate choice of regularization parameter, the lasso process drives most of the auxiliary channel coefficients to zero and constructs a model using a small subset of channels (usually a few to a dozen) with non-zero lasso coefficients. The list of channels with non-zero coefficients is ranked by the absolute values of their lasso coefficients, showing the relative contribution of each channel to the model, independent of sign. To visualize the model and its relation to the range, several summary plots are created, as shown in Figure~\ref{fig:modelplots}.

\subsection{Details of relevant channels}

 \begin{figure*}[ht]
 \begin{center}
 \begin{minipage}{0.8\linewidth}
  	\includegraphics[width=1.0\linewidth]{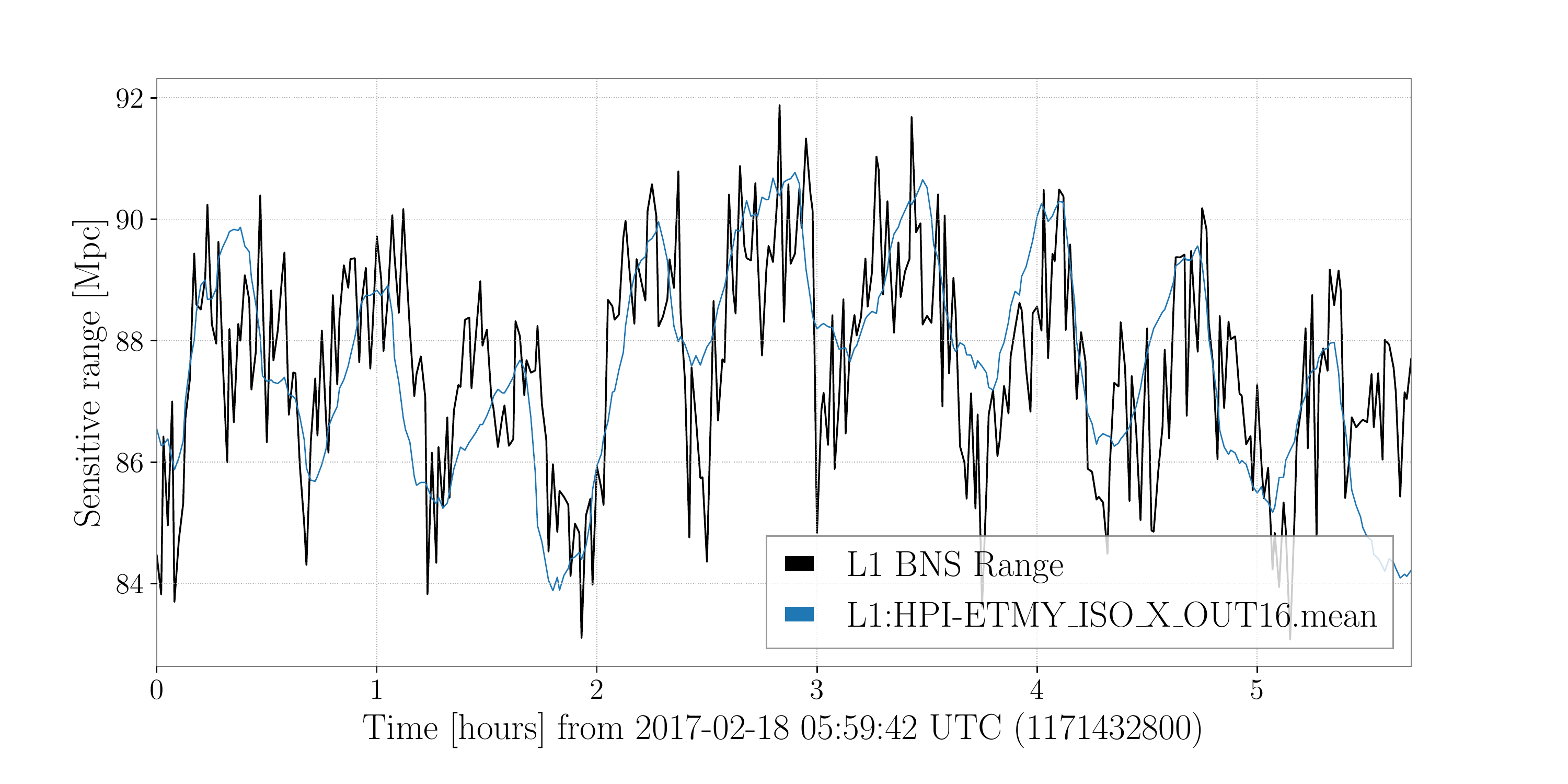}
 \end{minipage}
 \begin{minipage}{0.8\linewidth}
 	\includegraphics[width=1.0\linewidth]{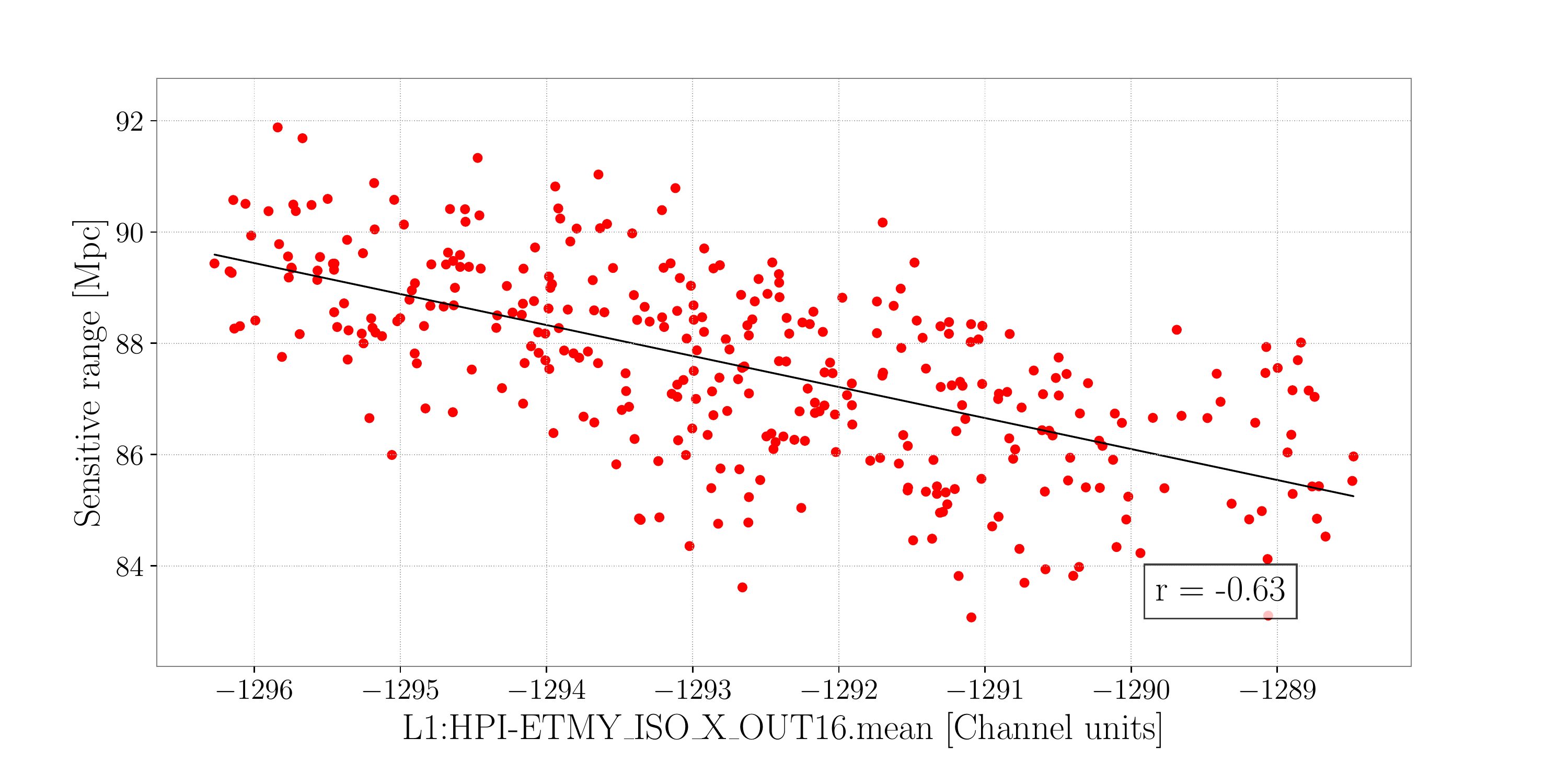}
 \end{minipage}
 \caption{Plots comparing fluctuations in an individual channel to those in the range. Above: Time series plot for the range channel and the auxiliary channel, both scaled. Below: Scatter plot of the same two channels in their original units, including the line of best fit and Pearson correlation coefficient found using ordinary linear regression. \label{fig:channelplots}} 
 \end{center}
 \end{figure*}

For each auxiliary channel used in the lasso model, additional plots and information are placed into an expandable tab in an accordion on the output webpage. To visualize how a given channel's fluctuations relate to the range's fluctuations, each tab contains subplots of the channel's unscaled time series and the range's unscaled time series, an overlay of the two time series scaled to Megaparsecs, and a scatter plot of the two channels' data (along with an assessment of linear correlation between that channel and the range), as shown in Figure~\ref{fig:channelplots}.   

\begin{figure*}[ht]
\begin{minipage}{0.5\textwidth}
	\includegraphics[width=1.0\linewidth]{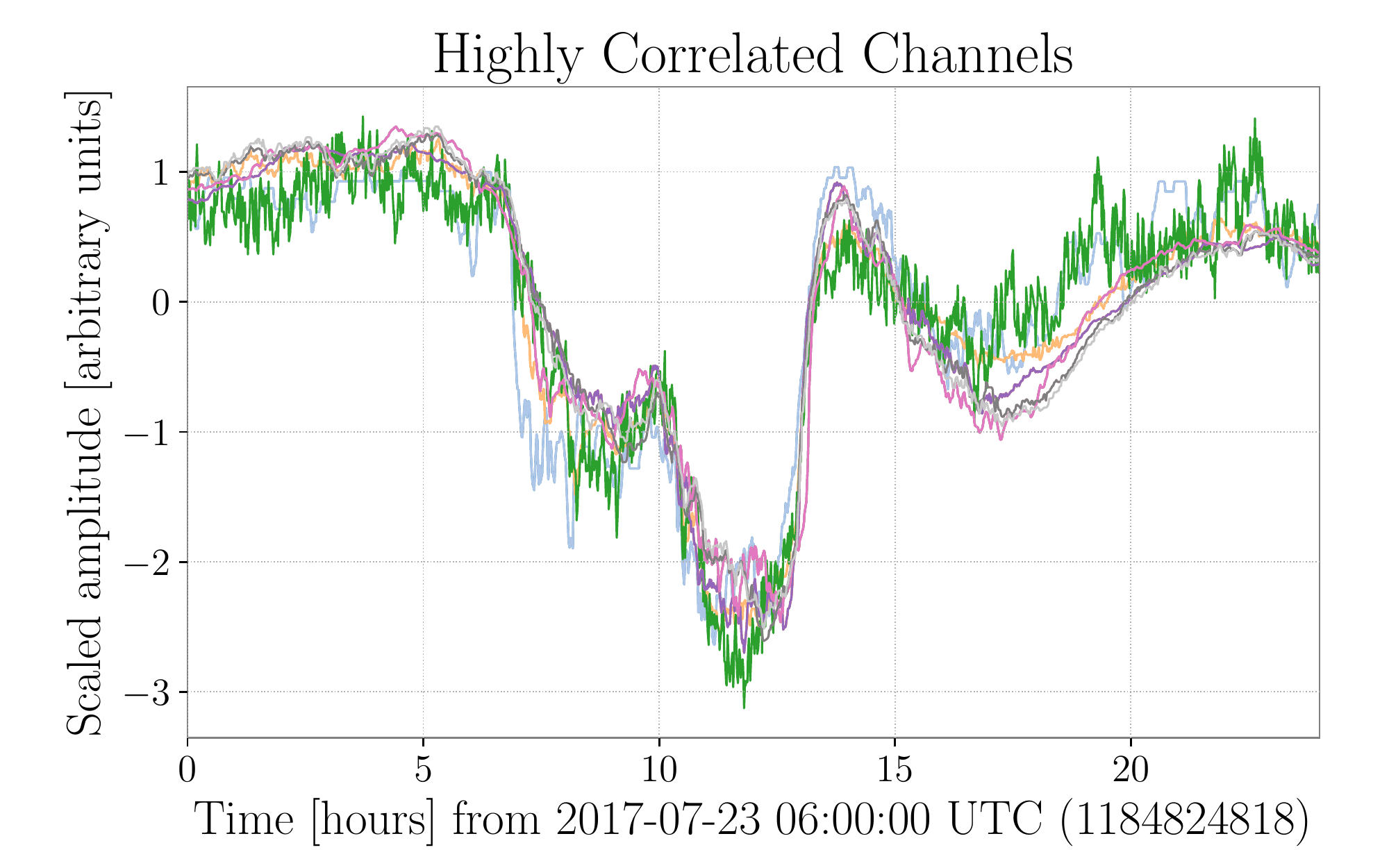}
\end{minipage}
\begin{minipage}{0.5\textwidth}
	\includegraphics[width=1.0\linewidth]{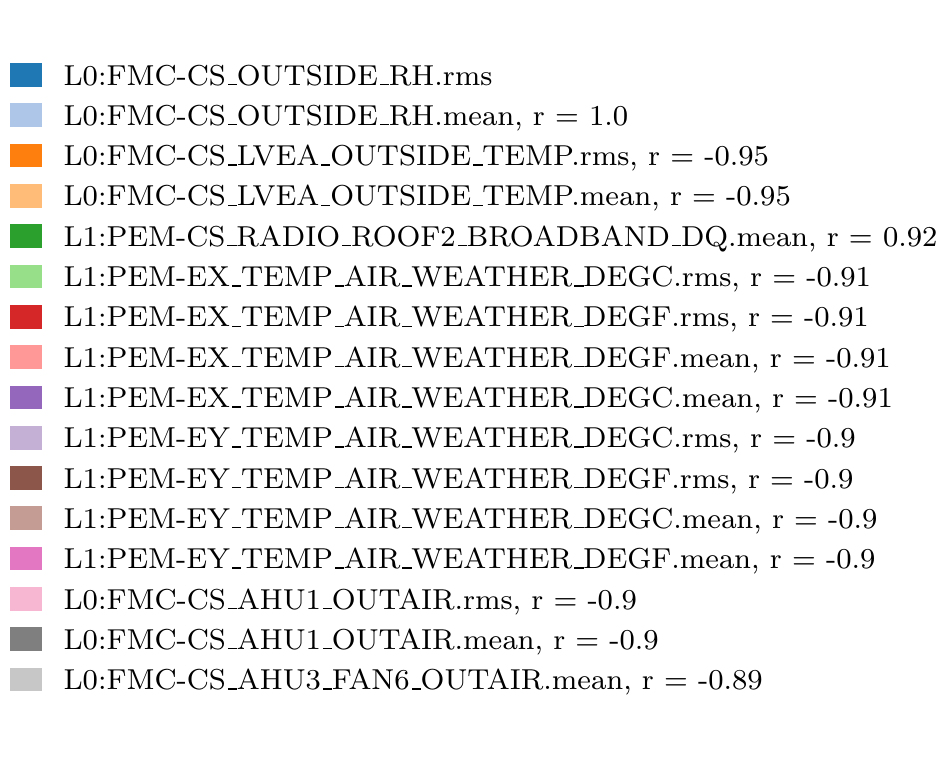}
\end{minipage}

 \caption{Relative humidity outside of the LIGO Livingston corner station, and a group of 12 other channels identified as closely correlated (Pearson coefficient absolute value greater than 0.9). Related channels include other measures of humidity, temperatures, fans, and a monitor of radio waves. (See \cite{Luomala2015EffectsOT} for an analysis of the effects of temperature and humidity on radio signal sensors.)\label{fig:cluster}} 
 \end{figure*}

Many of LIGO's auxiliary channels are themselves highly correlated, and thus a channel selected by lasso may actually be a representative of a large group of channels. Knowing which channels belong to this group can be very useful in diagnosing the physical mechanism driving range fluctuations. For this reason, for each channel in the lasso model, a cluster of channels whose Pearson linear correlation coefficient is above a given threshold are identified, listed, and plotted. An example cluster of channels is shown in Figure~\ref{fig:cluster}.

\subsection{Webpage output}

Finally, an HTML webpage is generated containing the input parameters, lasso model information and plots, text file lists of channels that were found to be flat or contain NaNs, and an accordion with tabs for all channels used in the model, with their individual suite of plots. The accordion's channel containers are ranked and given a corresponding color determined by the absolute value of their lasso coefficient. 
The webpage is published to the LIGO Data Grid and thus made available to all members of the LIGO Scientific Collaboration.

\section{Exemplary Results}

To be useful for gravitational-wave science, the algorithm should be able to robustly run on long stretches (hours to days) of data and identify key sets of auxiliary channels that relate to variations in the range.  To test its performance, we ran the algorithm on data from three representative situations: i) where a known single instrument malfunctioned, strongly impacting the range, ii) where a local effect, sensed by a group of channels, affected the range, and iii) a typical stretch of data for which the drivers of range are unknown. 

\subsection{Range corruption from a single source: The Livingston resistance temperature detector}

\begin{figure}[ht]
\begin{centering}
\begin{minipage}[ht]{1\textwidth}
	\includegraphics[width=1.0\textwidth]{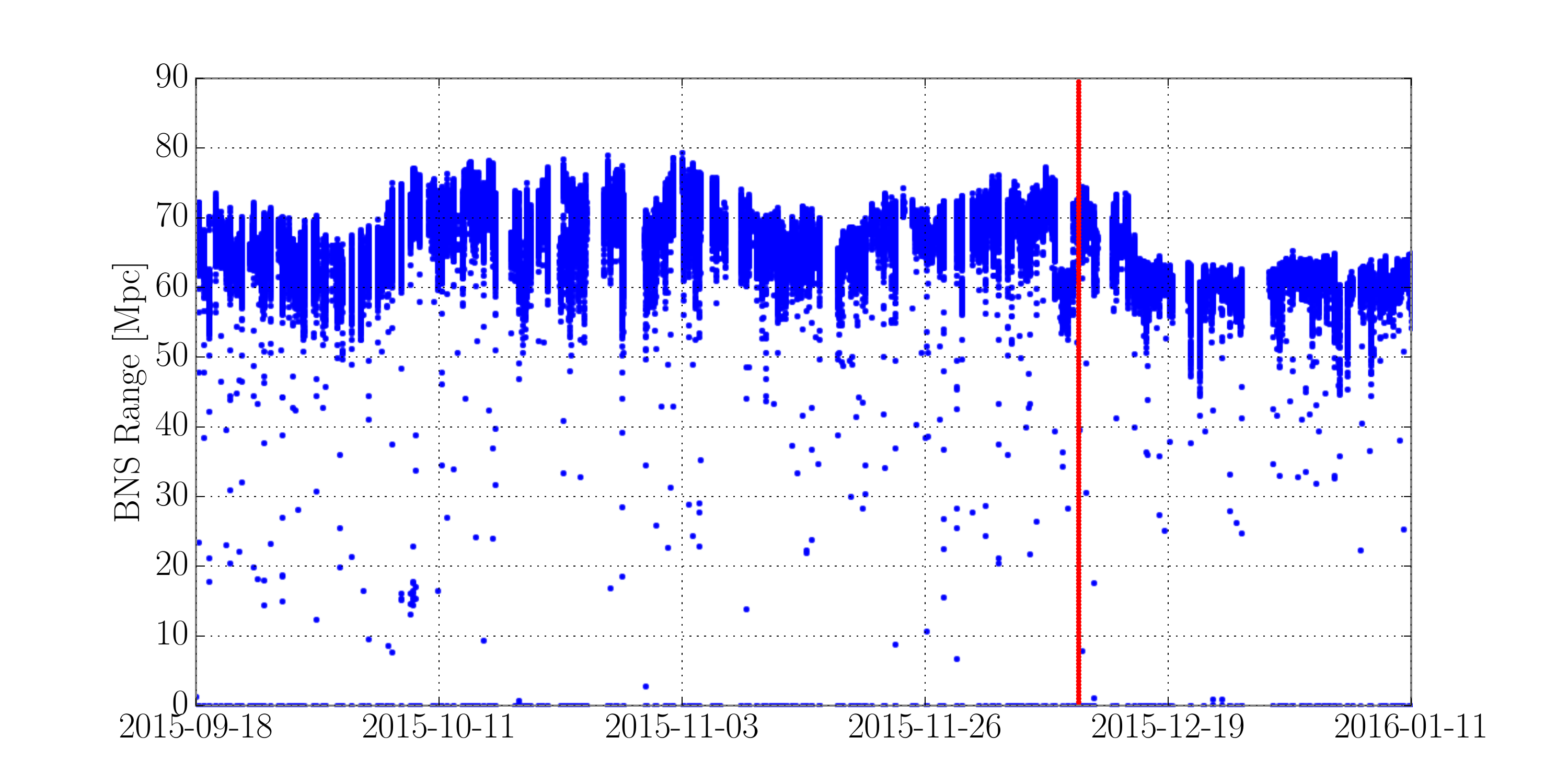}

	\includegraphics[width=1.0\textwidth]{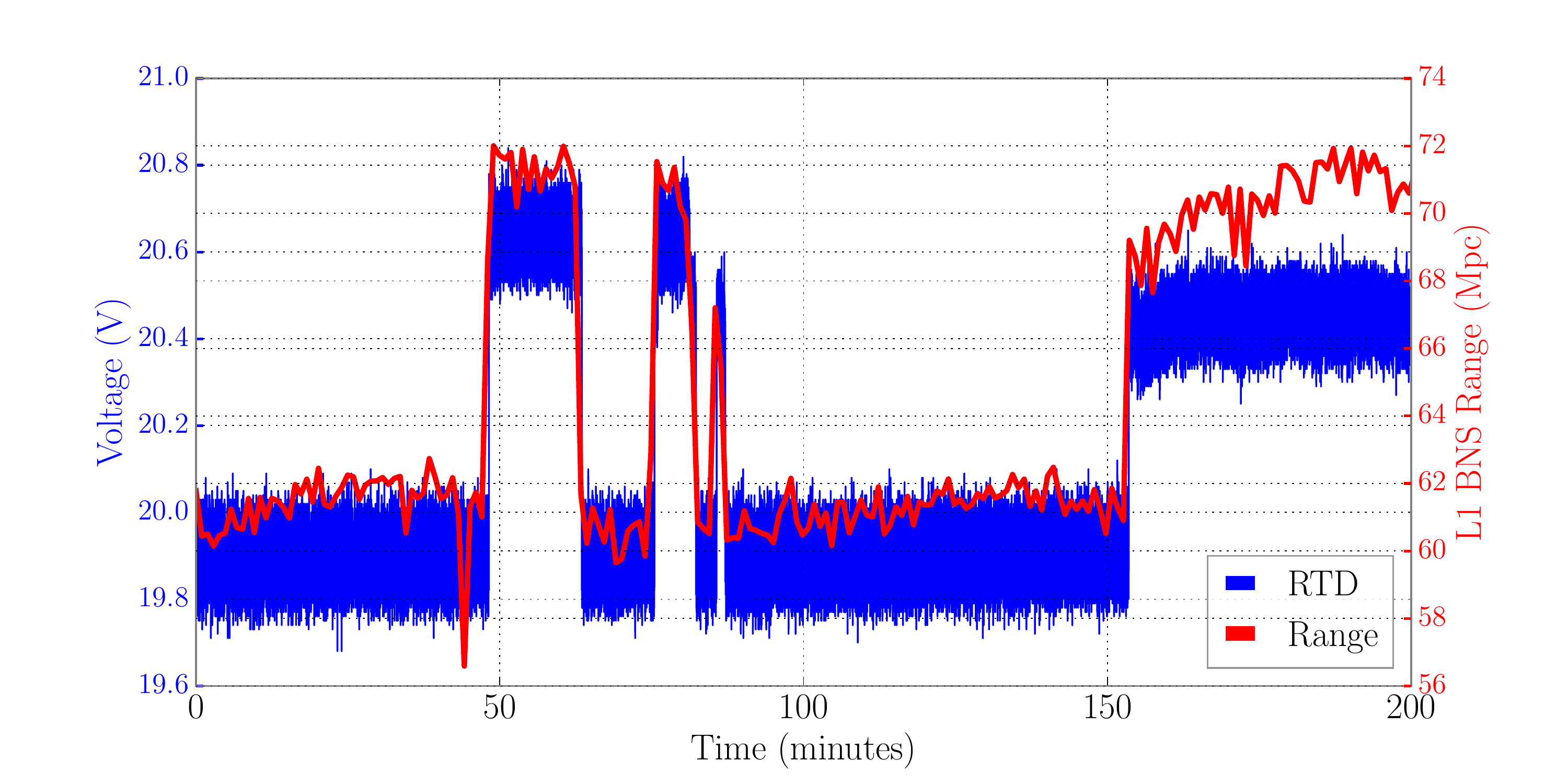}
\end{minipage}
\caption{Above: Binary neutron star range for LIGO Livingston for the O1 run~\cite{o1sensitivity}. Below: 200 minutes of Livingston BNS range data from December 11, 2015 13:29 UTC (right axis) and data from the malfunctioning resistance temperature detector (RTD) at the same time (left axis). The red vertical line in the upper plot indicates the start time of the lower plot.}\label{fig:o1range}
\end{centering}
\end{figure}

LIGO's first observing run started in September 2015 and ended in January 2016. Beginning in December 2015, until the end of the run, LIGO Livingston experienced a mysterious excess noise that caused its range to decrease significantly (from an average of 65 Mpc to an average of 60 Mpc) and sometimes exhibit sudden changes~\cite{o1sensitivity,SEI_RTD_Investigation,SEI_RTD_Investigation_Detchar}, as shown in Figure~\ref{fig:o1range}.  
After the observing run, instrument scientists on site determined that misbehaving electronics in Livingston's Y-end station were responsible for the excess noise. Specifically, the broken sensor was a resistance temperature detector (RTD). This sensor was used to measure the temperature of equipment in the vacuum enclosures and was not intended to influence the interferometer in any way. However, during these times the sensor exhibited excess noise that coupled through nearby equipment to reduce LIGO Livingston's range. 
Once the cause of this noise was discovered, it was, in retrospect, evident that the RTD sensor's noise fluctuations at times clearly resembled the variations in Livingston's range (right of Figure~\ref{fig:o1range}). Based on this experience, a design goal for our algorithm is that it be able to quickly and automatically identify any auxiliary channel, such as this one, that strongly correlates with changes in the range. 

\begin{figure}[ht]
\begin{center}
\begin{minipage}{0.8\textwidth}
	\includegraphics[width=1.0\textwidth]{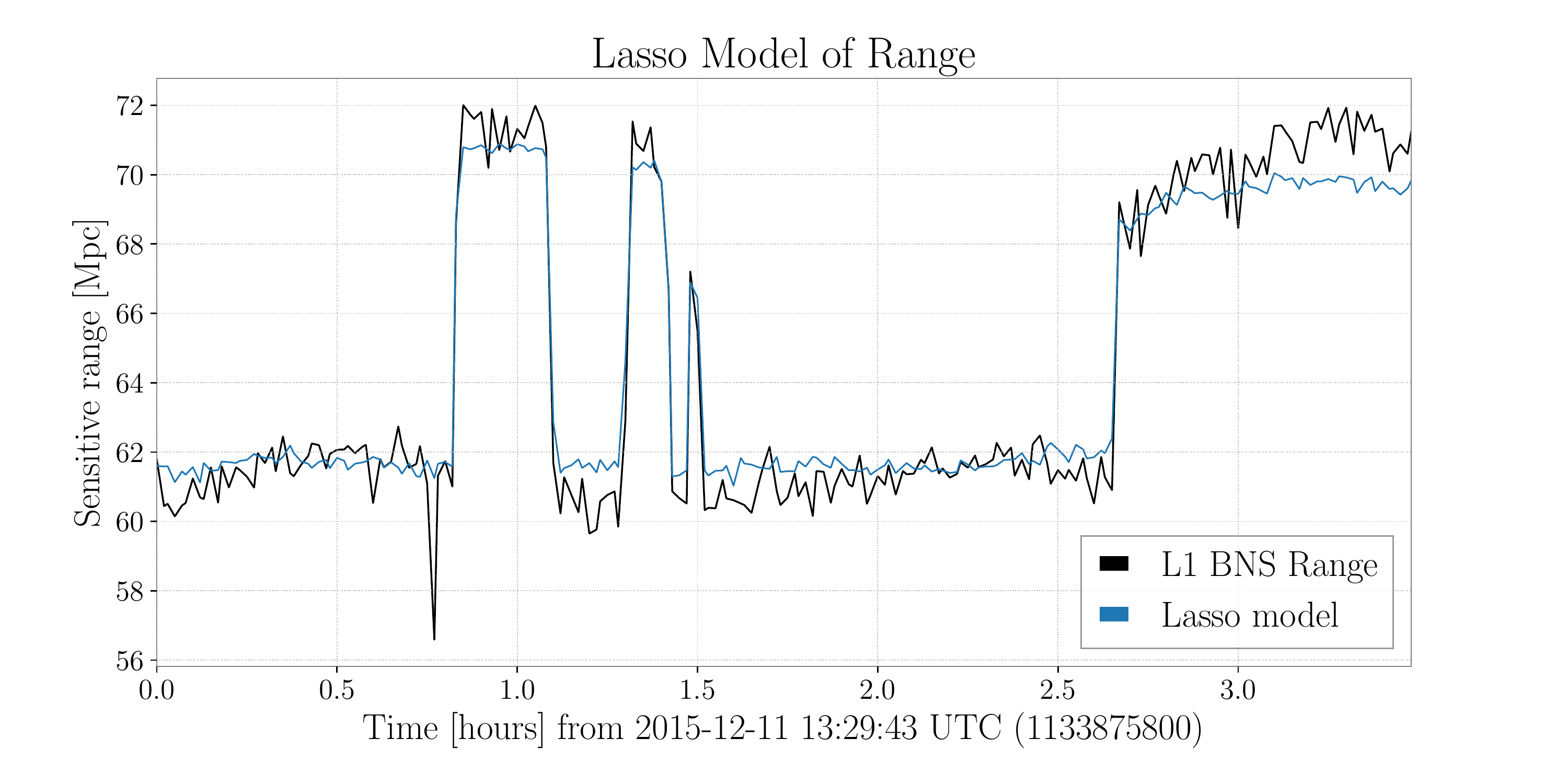}
\end{minipage}
\begin{minipage}{0.8\textwidth}
	\resizebox{\textwidth}{!}{
		\begin{tabular}{ | l | l | l | }
	    	\hline
			 &	Channel &	Lasso coefficient  \\ \hline
			1 &	L1:SEI-Y\_RTD\_STAGE1\_SENSORB.rms &	0.660721  \\ \hline
			2 &	L1:IOP-ASC\_Y\_TR\_B\_DEMOD\_PIT\_I\_OUT\_DQ.rms &	-0.163092 \\ \hline
			3 &	L1:CAL-PCALY\_TRANSMITTERMODULETEMPERATURE.mean &	0.042102  \\ \hline
			4 &	L1:SUS-SR2\_M1\_RMSIMON\_LF\_MON.rms &	-0.015628  \\ \hline
			5 &	L1:SEI-Y\_RTD\_STAGE1\_SENSORB.mean &	0.011090 \\ \hline
			6 &	L1:SUS-SR2\_M1\_RMSIMON\_LF\_MON.mean &	-0.000435  \\ \hline
		\end{tabular}}
\end{minipage}
\end{center}
\caption{Model for range data during the resistance temperature detector malfunction. Above: Binary neutron star range for LIGO Livingston between 2015-12-11 13:29:43 and 16:59:43 UTC and the lasso model. Below: Individual channel contributions to the lasso model. }\label{fig:rtd}
\end{figure}

Figure~\ref{fig:rtd} shows the lasso model results for a (more than three hour) period of time during which the RTD sensor was malfunctioning. Using the full set of possible minute-trend data from LIGO Livingston at this time (395,562 auxiliary channels, of which 67,360 were not flat) the algorithm identified the RTD sensor as its most significant channel and only used six channels in its model. Additionally, clustering identified no other closely related channels to the RTD sensor. Had this algorithm existed in 2015, it may have allowed LIGO scientists to identify and fix this source of excess noise one or two months earlier. 

\subsection{Range variations driven by a local effect: Livingston's Y-end station}

\begin{figure}[ht]
\begin{center}
\begin{minipage}{0.8\textwidth}
	\includegraphics[width=1.0\textwidth]{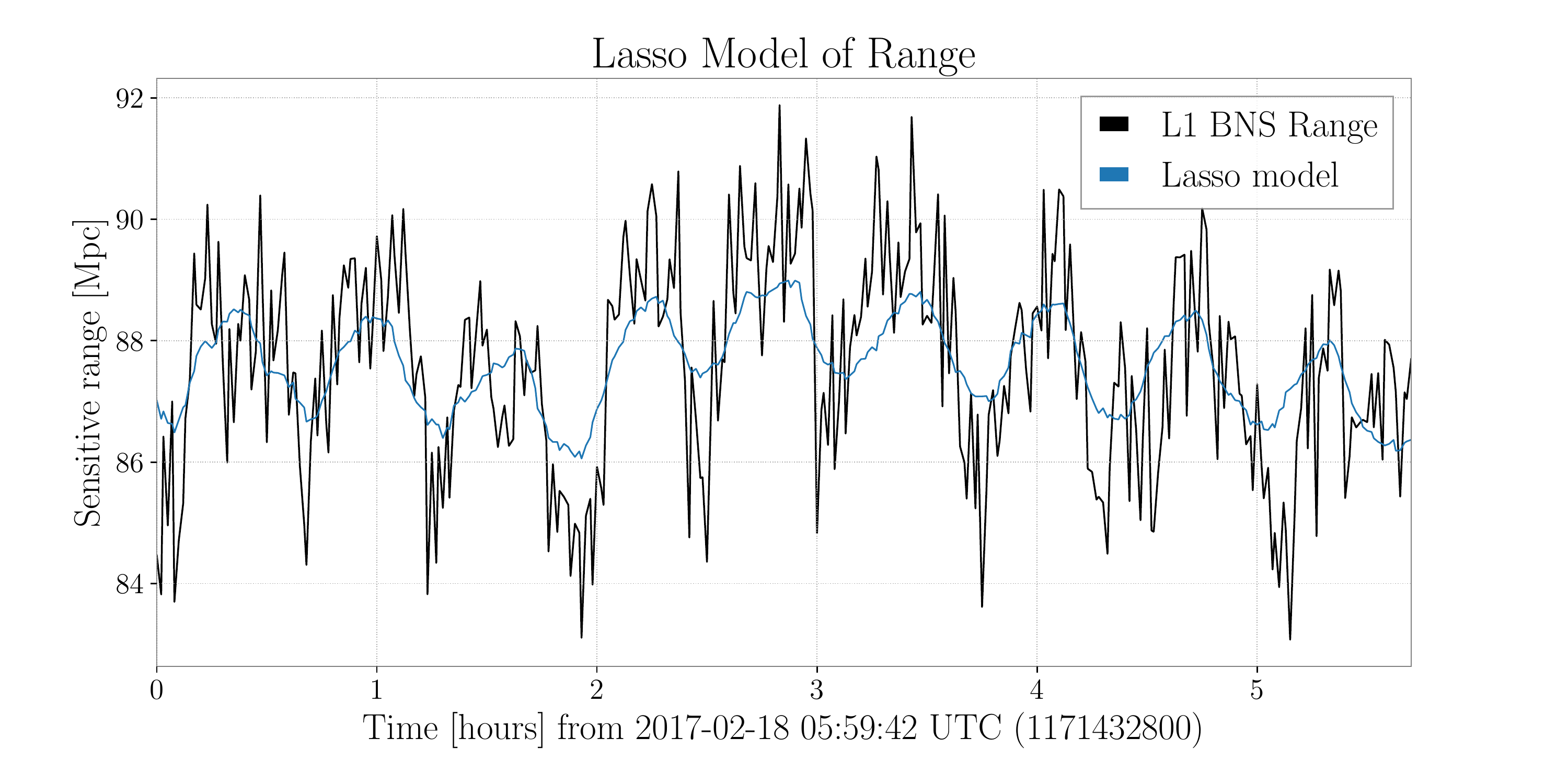}
\end{minipage}
\begin{minipage}{0.8\textwidth}
    \resizebox{\textwidth}{!}{
		\begin{tabular}{ | l | l | l |}
	    	\hline
	   	 	 &  Channel & Lasso coefficient \\ \hline
			1 &	L1:HPI-ETMY\_ISO\_X\_OUT16.mean &	-0.307973  \\ \hline
			2 &	L1:SUS-ETMY\_R0\_DAMP\_Y\_IN1\_DQ.rms &	-0.070306  \\ \hline
			3 &	L1:SYS-TIMING\_Y\_FO\_A\_PORT\_9\_SLAVE\_GENERIC\_PAYL... &	0.059067 \\ \hline
			4 &	L1:HPI-ETMY\_BLRMS\_Z\_30M.mean &	0.029452  \\ \hline
			5 &	L1:OAF-CAL\_DARM\_DQ.rms &	-0.027925  \\ \hline
			6 &	L1:ISI-BS\_ST1\_FFB\_BLRMS\_RY\_10\_30.mean &	-0.021117  \\ \hline
			7 &	L1:SUS-SR3\_M1\_VOLTMON\_T2\_MON.rms &	-0.019242  \\ \hline
			8 &	L1:HPI-ITMX\_BLRMS\_RX\_10\_30.rms &	-0.005955 \\ \hline
			9 &	L1:HPI-ETMY\_BLRMS\_Z\_30M.rms &	0.000236  \\ \hline
			10 &	L1:SUS-SR3\_M1\_VOLTMON\_T2\_MON.mean &	0.000153  \\ \hline
		\end{tabular}}
\end{minipage}
\end{center}
\caption{Model for range data during times with range variations driven by sources local to the Livingston Y-end station. Above: Binary neutron star range for LIGO Livingston between 2017-02-18 05:59:42 and 11:42:42 UTC and the lasso model. Below: Individual channel contributions to the lasso model.\label{fig:ETMY}} 
\end{figure}

Early in LIGO's second observing run, Livingston's BNS range exhibited a strong periodic variation, often 5\,Mpc peak-to-peak, corresponding to variations of the amplitude of a rounded bump of noise between 60 and 70\,Hz in the detector's strain spectrum. Various studies associated the cause of the noise variations to be local to the Y-end station~\cite{etmyalogjosh,etmyalogrobert}.  
Figure~\ref{fig:ETMY} shows a 5 hour stretch of LIGO Livingston data from February 18, 2017, when this noise was particularly apparent. The strong 35-minute periodicity seen in the range is related to the cycle time of the air handling units used to stabilize the temperature of the Y-end station.

Figure~\ref{fig:ETMY} also shows the lasso model results using 415,490 auxiliary channels (69,583 non-flat channels) that were recorded at this time. The dominant contributor to the model is motion of the vacuum tank enclosing the Y-end mirror in the direction transverse to the beam, which is shown in Figure~\ref{fig:channelplots}. The four channels with the largest coefficients are all related to seismic motion in the Y-end station. Additionally, correlation analysis shows that each of these channels is part of a larger cluster of Y-end station seismic channels exhibiting similar fluctuations. 

Because many observables in the Y-end station are similarly driven by the air handling unit, this model is more complex and less interpretable than the RTD sensor example above. However, it does correctly identify the general location of the noise driver. Several other periods in the days around this example exhibited similar range fluctuations. The lasso models for these days also contained many channels from the Y-end station, supporting this as a sustained effect.

\subsection{Range variations with unknown noise sources: a typical day at Livingston}

\begin{figure}[ht]
\begin{center}
\begin{minipage}{0.8\textwidth}
	\includegraphics[width=1.0\textwidth]{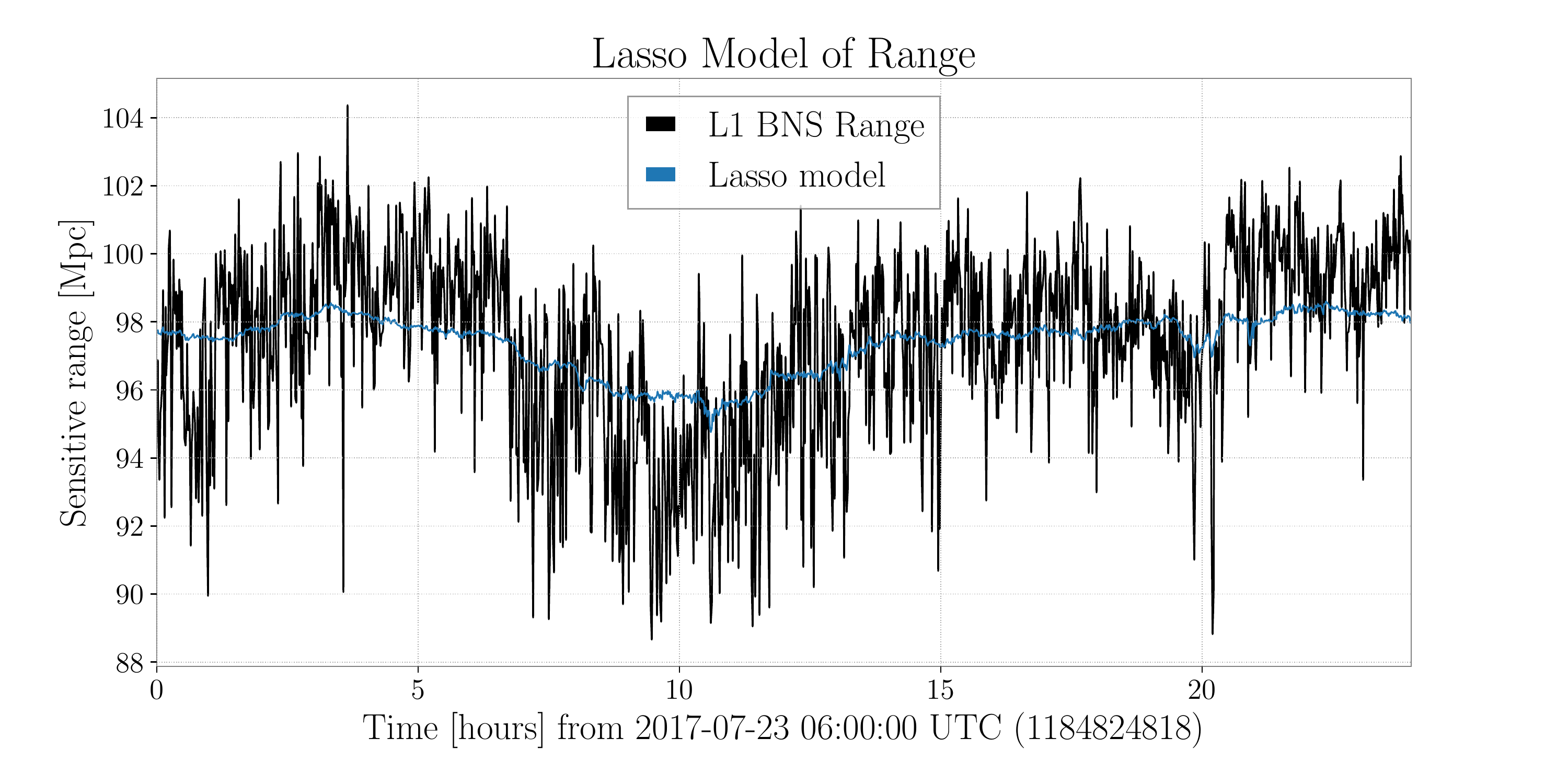}
\end{minipage}
\begin{minipage}{0.8\textwidth}
    \resizebox{\textwidth}{!}{
    	\begin{tabular}{ | l | l | l |}
	   		\hline
			 &	Channel &	Lasso coefficient \\ \hline
			1 &	L0:FMC-CS\_OUTSIDE\_RH.rms &	0.084270 \\ \hline
			2 &	L1:SUS-SRM\_M2\_NOISEMON\_LL\_OUT\_DQ.rms &	-0.063073 \\ \hline
			3 &	L1:PSL-PWR\_HPL\_DC\_LP\_OUTPUT.rms &	-0.055047 \\ \hline
			4 &	L1:ISI-ETMY\_ST2\_SCSUM\_CPS\_RX\_IN\_DQ.rms &	-0.038619 \\ \hline
			5 &	L1:ISI-ETMY\_ST1\_GNDSTSINF\_C\_Z\_INMON.rms &	-0.034891 \\ \hline
			6 &	L1:PSL-PWR\_NPRO\_OUT16.rms &	-0.033688 \\ \hline
			7 &	L1:SUS-MC2\_M1\_DAMP\_P\_IN1\_DQ.rms &	0.027549 \\ \hline
			8 &	L1:SUS-RM1\_M1\_WD\_OSEMAC\_LL\_RMSMON.mean &	-0.021243 \\ \hline
			9 &	L1:PSL-PWR\_NPRO\_INMON.mean &	0.013105 \\ \hline
			10 &	L1:IOP-PSL0\_MADC2\_EPICS\_CH29.mean &	0.011312 \\ \hline
			11 &	L1:SUS-RM1\_M1\_VOLTMON\_LL\_DQ.mean &	-0.009482 \\ \hline
			12 &	L1:PSL-PWR\_NPRO\_OUT16.mean &	-0.000958 \\ \hline
			13 &	L1:ISI-ETMY\_ST1\_GNDSTSINF\_C\_Z\_OUTPUT.rms &	-0.000449 \\ \hline
		\end{tabular}}
\end{minipage}
\end{center}
\caption{Lasso model for range data during a typical LIGO Livingston day. Above: Binary neutron star range for LIGO Livingston between 2017-07-23 06:00:00 and 2017-07-23 06:00:00 UTC and the lasso model. Below: Individual channel contributions to the lasso model. \label{fig:24h}} 
\end{figure}

The previous two examples demonstrate that in situations where the drivers of the range are strong and understood, our algorithm produces lasso models that identify channels that are related to the cause. Often, however, the range is driven by multiple unknown noise sources interacting in subtle ways, presenting a much greater challenge. Figure~\ref{fig:24h} shows the range from a relatively stable day at LIGO Livingston taken from LIGO's second observing run. The BNS range of the detector drifted slowly over the course of a full day without any of the sharp jumps or obvious periodicity dominating the other two examples. 

\begin{figure*}[ht]
\begin{minipage}{0.6\textwidth}
	\includegraphics[width=1.0\linewidth]{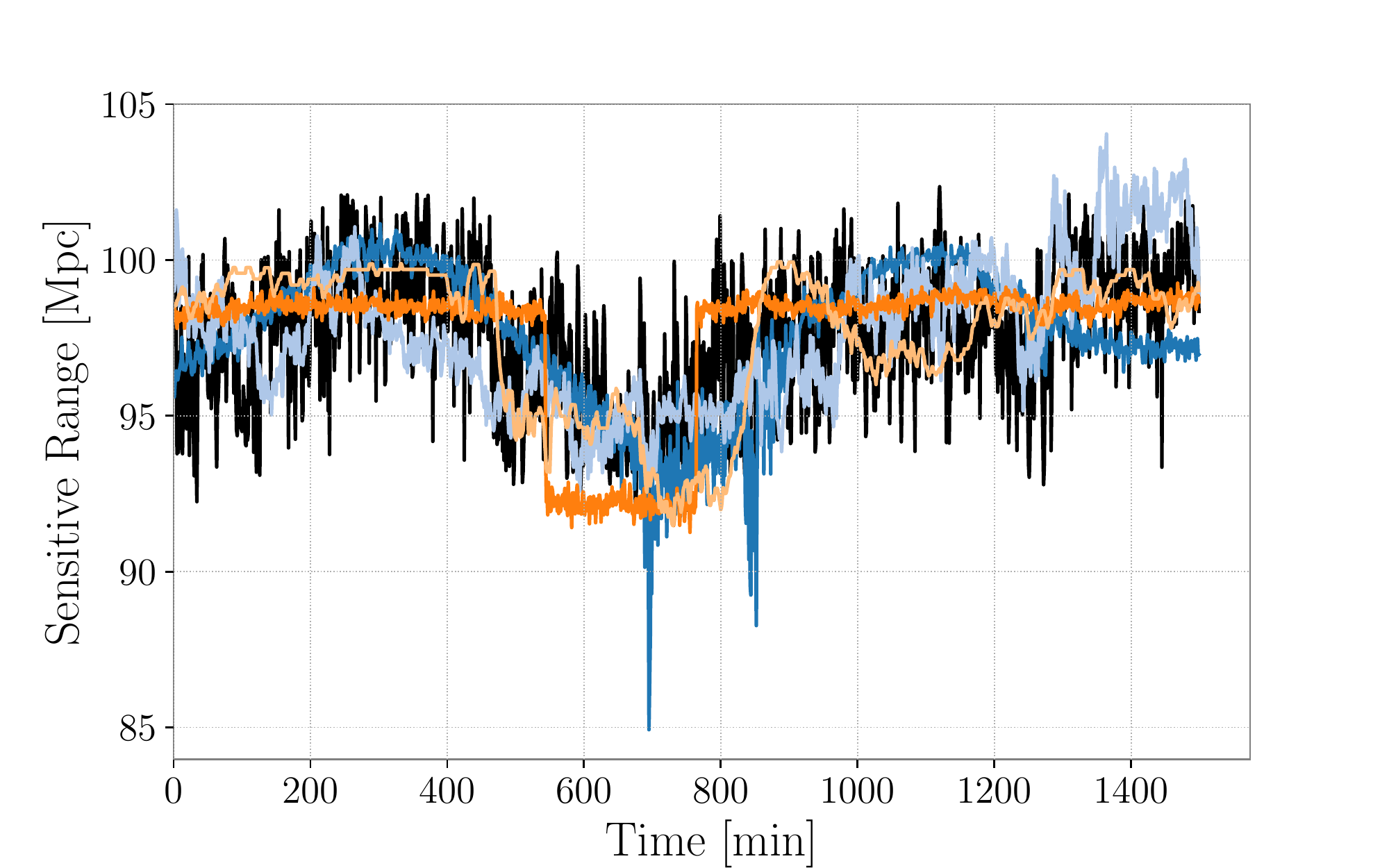}
\end{minipage}
\begin{minipage}{0.4\textwidth}
	\includegraphics[width=1.0\linewidth]{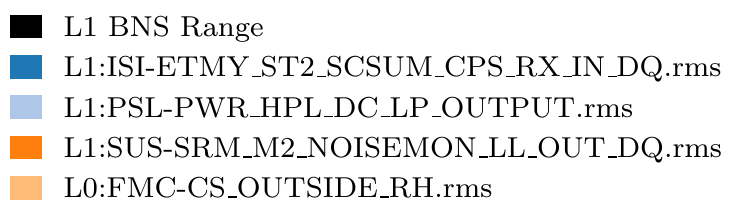}
\end{minipage}
\caption{BNS range for the day shown in Figure~\ref{fig:24h} and, overlaid, the time series of the top four auxiliary channels in the lasso model, each scaled and multiplied by the sign of their correlation coefficient. \label{fig:overlay24h}}
\end{figure*}

The lasso model for this period is also shown in Figure~\ref{fig:24h}. Of the three examples, the leading coefficient in this model is the lowest. The first four leading contributors to the model are shown scaled and overlaid with the range in Figure~\ref{fig:overlay24h}. The channel with the largest coefficient measures the relative humidity outside of the LIGO corner station. It is highly correlated with several measures of humidity and temperature across the Livingston site, as shown in Figure~\ref{fig:cluster}. 
The channel with the second highest coefficient is designed to measure the electronics noise in a displacement sensor on the pendulum suspending the signal recycling mirror. Other sensors from the same pendulum do not exhibit the same behavior, so it is likely that this channel is not functioning properly. This channel exhibits a sharp step that is coincident with a drop in the range, so it helps the model but is not physically related.
Further channels used by the model include the power of the laser and tidally-driven ground motion, each of which exhibit a quite similar shape to the range variations, and could, conceivably, be related to noise drivers of the range, though further follow-up is needed to understand the potential noise coupling mechanisms. 

This example reveals some of the limitations of this algorithm when there is not a single clear source of noise. Lasso identifies a diverse collection of channels that each exhibit some of the characteristics of the range fluctuations, but the model is not entirely convincing. For cases like this, the tool does not provide answers so much as potential starting points for deeper investigations. 
One way to identify promising channels to follow up is to compare the results of the algorithm using different sub-segments of the same day or different values for the alpha parameter.
For this example, dividing the day into different time segments and varying alpha led to mixed results, with only a subset of channels appearing consistently.

\subsection{Choice of alpha}

\begin{figure*}[ht]
\begin{center}
\includegraphics[width=0.9\linewidth]{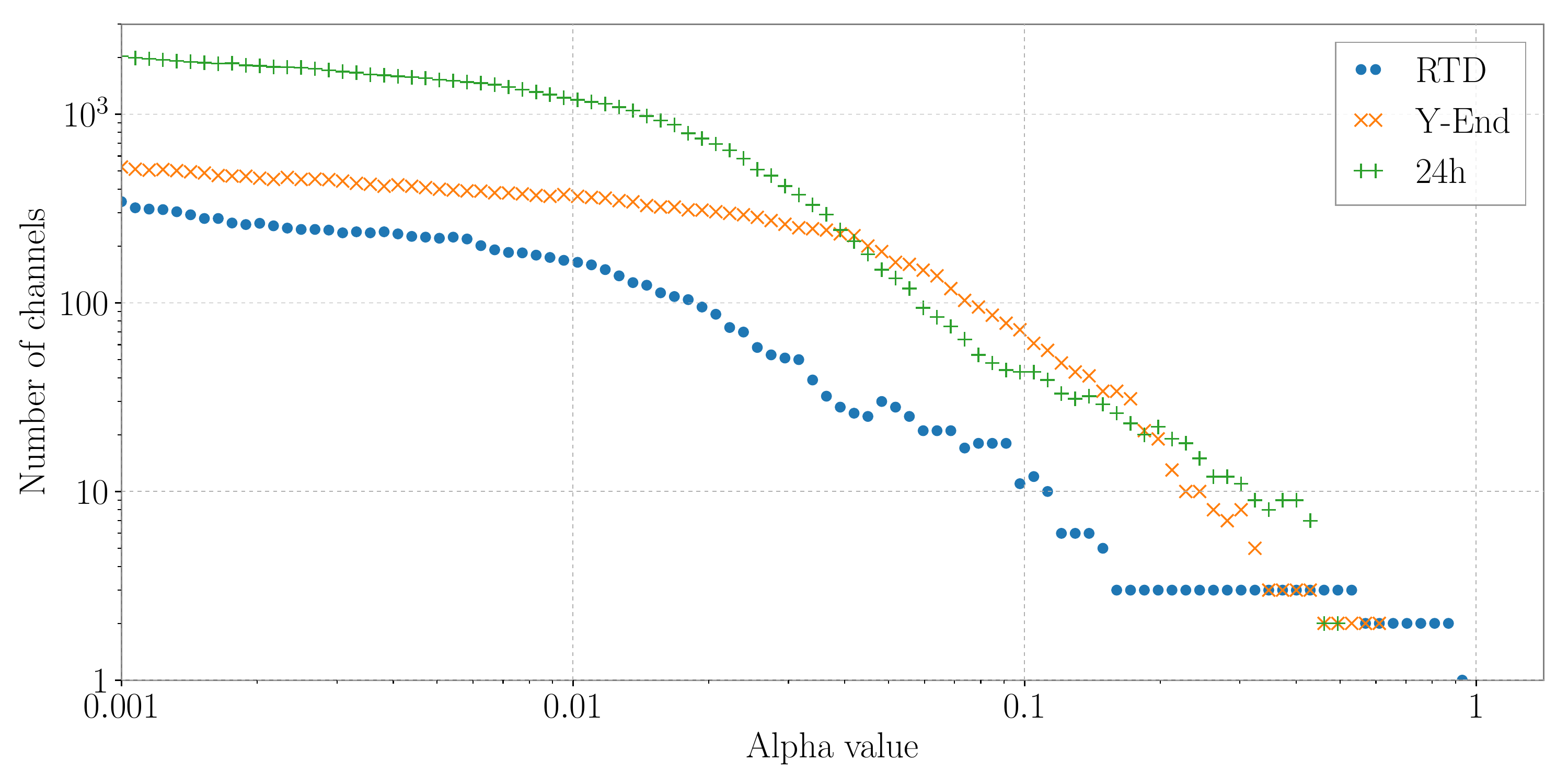}
\caption{Number of channels used in the lasso model versus the value of alpha for the times of the three examples discussed above. The actual alpha values used in the examples were 0.12, 0.22, and 0.25, for the resistance temperature detector (RTD), Y-end, and 24-hour cases, respectively. For an alpha of zero (not shown), the lasso model reduces to ordinary least squares, and all non-flat channels (about 70,000) are used. \label{fig:alpha-nchans}} 
\end{center}
\end{figure*}

For the examples above, we adjusted the alpha values to obtain interpretable models. Figure~\ref{fig:alpha-nchans} shows how the number of channels in the model varies with the chosen alpha value. The RTD model uses fewer channels at all values of alpha and converges to three channels for alpha greater than 0.15. This behavior is likely due to the fact that the primary driver of the range variation is one broken sensor, which dominates the model. The 24-hour model uses many more channels, still using more than 10 channels even with an alpha value of 0.3. This points to a more intricate combination of potential sources of noise. 
The Y-end station model is more complex than the RTD model, but at low alphas requires significantly fewer channels than the much more complicated 24-hour model. 
% The situation where we have a localized source in the Y-end station is mostly in between these two cases. 
These examples illustrate the variability and complexity of the lasso models when applied to data with different lengths of time and strengths of noise coupling. 

\subsection{Computational costs}
These results were generated on the LIGO Data Grid. They were each run on the LIGO Livingston computer cluster's head node, named \textit{ldas-pcdev5}, which has 2.2GHz Xeon E5-2698v4 processors and 512GB of RAM. With multiprocessing enabled (up to 20 CPUs), the wall clock time for the runs were 0:21:44.9, 0:44:46.6, and 2:25:31.4 for the RTD (3 hours of data), ETMY (5 hours of data), and 24-hour (24 hours of data) results, respectively. The limiting factor in each case was data retrieval. 

\section{Conclusions}

In this paper we have described an algorithm that uses lasso regression to construct interpretable models of variations in LIGO's range using auxiliary channels and demonstrated its application to three periods of LIGO data. Work is currently underway to automate this algorithm for daily running and display on the LIGO Summary Pages~\cite{gwsumm, gwpy}, with the goal of using the results to quickly identify and resolve problematic instrumental couplings. For this purpose, we are investigating ways to automate the choice of alpha so that the algorithm can reliably produce useful results. Promising methods for making this choice include cross-validation and information criteria~\cite{james2013introduction}.

As with any correlation method, these results do not imply causality. It is possible, or even likely, that because of noise in the recording process or frequency-dependent coupling, that the auxiliary channel with the very highest correlation is not the one that is most closely related to the noise source. It will always be necessary to follow up the results of this algorithm with physical calculations or experiments to test for bona fide couplings. Toward this end, we have begun investigating interesting correlations identified by this method in past LIGO data, but that work is beyond the scope of this paper. A closely related tool that will inform future investigation into the physical noise mechanisms is the \textit{noise budget}, which measures the estimated spectral amplitude with which various types of known noise couple to the detector output~\cite{o1sensitivity,smith2006linear,ranathesis,vajente2008analysis}.

This algorithm is a relatively simple application of statistical regression techniques to LIGO data. It does not, for example, take into account time-dependent coupling, nonlinearity or phase shifts, issues that are known to play a role in gravitational-wave detectors. We are exploring other techniques, such as distance correlation~\cite{székely2007,richards2017distance} and Fourier methods, that show promise for incorporating more rich couplings. 

\ack{The authors are grateful to their colleagues in the LIGO Scientific Collaboration for stimulating discussions and for review of this manuscript, especially Gabriele Vajente for helpful comments in the internal review process. The authors would also like to thank the CQG referees for insightful suggestions in their reviews, which have led to improvements in the manuscript and algorithm. MW and JS are pleased to acknowledge the support of Dan O. Black and family. This research was supported by National Science Foundation grants NSF PHY-1255650, AST-1559694, PHY-1708035, and PHY-1807069. Computations in this paper were performed on the LIGO Data Grid.}

\appendix
\section*{Appendix - Channel Information}
\setcounter{section}{1}

The LIGO channels shown in this paper have long and complicated names, using a series of abbreviations to specify the exact location, subsystem, and type of sensors being recorded, according to a convention described in~\cite{channelnaming}. We have shown their full names so this analysis could be reproduced using the same unique channels, and here give brief descriptions of the channels.

\subsection{Environmental Monitors}

\begin{itemize}
\item L1:SEI-Y\_RTD\_STAGE1\_SENSORB
Resistance Temperature Detector associated with the seismic isolation system at the the Y-end station

\item L0:FMC-CS\_OUTSIDE\_RH
Relative humidity outside the corner station

\item L0:FMC-CS\_LVEA\_OUTSIDE\_TEMP
Temperature outside the corner station

\item L1:PEM-CS\_RADIO\_ROOF2\_BROADBAND\_DQ
Broadband radio signal at the corner station

\item L1:PEM-E\{X,Y\}\_TEMP\_AIR\_WEATHER\_DEG\{C,F\}
%L1:PEM-EX\_TEMP\_AIR\_WEATHER\_DEGF
%L1:PEM-EY\_TEMP\_AIR\_WEATHER\_DEGC
%L1:PEM-EY\_TEMP\_AIR\_WEATHER\_DEGF
Air temperature at the X- or Y-end station in Celsius or Fahrenheit
\end{itemize}

\subsection{Pre-stabilized laser system (PSL)}

\begin{itemize}
\item L1:PSL-PWR\_HPL\_DC\_LP\_OUTPUT
Pre-stabilized high-power laser power	

\item L1:PSL-PWR\_NPRO\_\{INMON,OUT16\}
Pre-stabilized laser power out of the non-planar ring oscillator

\item L1:IOP-PSL0\_MADC2\_EPICS\_CH29
Pre-stabilized laser power
\end{itemize}

\subsection{Seismic motion measured at different points in the seismic isolation systems}

Hydraulic external pre-isolation of the vacuum chambers (HPI):
\begin{itemize}
\item L1:HPI-ETMY\_ISO\_X\_OUT16
Y-end HPI in the X direction

\item L1:HPI-ETMY\_BLRMS\_Z\_30M
Y-end seismic motion in the Z direction band limited up to 30 mHz

\item L1:HPI-ITMX\_BLRMS\_RX\_10\_30
Seismic motion at the input to the X-arm, band limited 10 to 30 Hz, in the direction rotating around the X axis
\end{itemize}

Internal Seismic Isolation inside vacuum chambers (ISI):
\begin{itemize}
\item L1:ISI-BS\_ST1\_FFB\_BLRMS\_RY\_10\_30
Seismic isolation of the beam splitter, band limited from 10 to 30 Hz, in the direction rotating around the Y axis

\item L1:ISI-ETMY\_ST2\_SCSUM\_CPS\_RX\_IN\_DQ
Capacitive position sensor, second stage of internal seismic isolation of the Y-end test mass, in the direction rotating around the X axis

\item L1:ISI-ETMY\_ST1\_GNDSTSINF\_C\_Z\_\{INMON,OUTPUT\}
Y-end seismic isolation in the Z direction %(INMON and OUTPUT are the same signal before and after filters)
\end{itemize}

\subsection{Suspension sensors}

\begin{itemize}
\item L1:SUS-ETMY\_R0\_DAMP\_Y\_IN1\_DQ
Displacement at the top stage of the Y-end reaction mass suspension in the yaw direction

\item L1:SUS-SRM\_M2\_NOISEMON\_LL\_OUT\_DQ
Lower left displacement sensor on the second stage of the signal recycling mirror

\item L1:SUS-SR2\_M1\_RMSIMON\_LF\_MON
Left front displacement sensor of the first stage of one of the mirror suspensions in the signal recycling cavity

\item L1:SUS-SR3\_M1\_VOLTMON\_T2\_MON
Voltage monitor in one of the actuators on the top stage of one of the mirror suspensions in the signal recycling cavity

\item L1:SUS-MC2\_M1\_DAMP\_P\_IN1\_DQ
Top stage pitch motion of the top stage of one of the mirror suspensions in the input mode cleaner cavity

\item L1:SUS-RM1\_M1\_WD\_OSEMAC\_LL\_RMSMON
Top stage lower left displacement sensor of a suspended steering mirror 

\item L1:SUS-RM1\_M1\_VOLTMON\_LL\_DQ
Voltage monitor of the lower left actuator on a suspended steering mirror 
\end{itemize}

\subsection{Other}
\begin{itemize}
\item L1:SYS-TIMING\_Y\_FO\_A\_PORT\_9\_SLAVE\_GENERIC\_PAYLOAD\_9
Monitor of the Y-end timing system

\item L1:OAF-CAL\_DARM\_DQ Calibrated differential length of the long arm cavities

\item L1:IOP-ASC\_Y\_TR\_B\_DEMOD\_PIT\_I\_OUT\_DQ
Pitch alignment signal sensed on the transmitted light from the y-arm 

\item L1:CAL-PCALY\_TRANSMITTERMODULETEMPERATURE
Transmitter module temperature in the photon calibrator system in the Y arm
\end{itemize}

%%%%%%%%%%%%%%%%%%%%%%%%%%%%%%%%%%%%%%%%%%%%%%%%%%%%%%%%%%%%%%%%%%%%%%%%%%%%%%%
\section*{References}
%%%%%%%%%%%%%%%%%%%%%%%%%%%%%%%%%%%%%%%%%%%%%%%%%%%%%%%%%%%%%%%%%%%%%%%%%%%%%%%
\bibliographystyle{iopart_num}
\bibliography{references}

\providecommand{\newblock}{}
\begin{thebibliography}{10}
\expandafter\ifx\csname url\endcsname\relax
  \def\url#1{{\tt #1}}\fi
\expandafter\ifx\csname urlprefix\endcsname\relax\def\urlprefix{URL }\fi
\providecommand{\eprint}[2][]{\url{#2}}

\bibitem{aligo}
Aasi J {\em et~al.\/} 2015 {\em Classical and Quantum Gravity\/} {\bf 32}
  074001 \urlprefix\url{https://doi.org/10.1088/0264-9381/32/7/074001}

\bibitem{gw150914}
Abbott B~P {\em et~al.\/} (LIGO Scientific Collaboration and Virgo
  Collaboration) 2016 {\em Phys. Rev. Lett.\/} {\bf 116}(6) 061102
  \urlprefix\url{https://link.aps.org/doi/10.1103/PhysRevLett.116.061102}

\bibitem{virgo2014}
Acernese F {\em et~al.\/} 2015 {\em Classical and Quantum Gravity\/} {\bf 32}
  024001 \urlprefix\url{http://stacks.iop.org/0264-9381/32/i=2/a=024001}

\bibitem{o1bbh}
{Abbott et al\ (The LIGO Scientific Collaboration and the Virgo Collaboration)}
  B 2016 {\em Phys. Rev. X\/} {\bf 6} 041015 (\textit{Preprint}
  \eprint{1606.04856})
  \urlprefix\url{https://journals.aps.org/prx/abstract/10.1103/PhysRevX.6.041015}

\bibitem{gw151226}
Abbott B~P {\em et~al.\/} (LIGO Scientific Collaboration and Virgo
  Collaboration) 2016 {\em Phys. Rev. Lett.\/} {\bf 116}(24) 241103
  \urlprefix\url{https://link.aps.org/doi/10.1103/PhysRevLett.116.241103}

\bibitem{gw170104}
Abbott B~P {\em et~al.\/} (LIGO Scientific and Virgo Collaboration) 2017 {\em
  Phys. Rev. Lett.\/} {\bf 118}(22) 221101
  \urlprefix\url{https://link.aps.org/doi/10.1103/PhysRevLett.118.221101}

\bibitem{gw170608}
Abbott B~P, others (LIGO Scientific~Collaboration and Collaboration) V 2017
  {\em The Astrophysical Journal Letters\/} {\bf 851} L35
  \urlprefix\url{http://stacks.iop.org/2041-8205/851/i=2/a=L35}

\bibitem{gw170814}
Abbott B~P {\em et~al.\/} (LIGO Scientific Collaboration and Virgo
  Collaboration) 2017 {\em Phys. Rev. Lett.\/} {\bf 119}(14) 141101
  \urlprefix\url{https://link.aps.org/doi/10.1103/PhysRevLett.119.141101}

\bibitem{gw170817}
Abbott B~P {\em et~al.\/} (LIGO Scientific Collaboration and Virgo
  Collaboration) 2017 {\em Phys. Rev. Lett.\/} {\bf 119}(16) 161101
  \urlprefix\url{https://link.aps.org/doi/10.1103/PhysRevLett.119.161101}

\bibitem{gw170817mma}
{The LIGO Scientific Collaboration}, {the Virgo Collaboration}, {Abbott} B~P,
  {Abbott} R, {Abbott} T~D, {Acernese} F, {Ackley} K, {Adams} C, {Adams} T,
  {Addesso} P and et~al 2017 {\em Ap. J. Lett.\/}
  \urlprefix\url{https://doi.org/10.3847/2041-8213/aa91c9}

\bibitem{Finn1993}
Finn L~S and Chernoff D~F 1993 {\em Phys. Rev. D\/} {\bf 47}(6) 2198--2219
  \urlprefix\url{https://link.aps.org/doi/10.1103/PhysRevD.47.2198}

\bibitem{observingscenarios}
Abbott B~P {\em et~al.\/} 2016 {\em Living Reviews in Relativity\/} {\bf 19} 1
  ISSN 1433-8351 \urlprefix\url{https://doi.org/10.1007/lrr-2016-1}

\bibitem{VajenteNonNA}
Vajente G 2015 At the border of commissioning and detchar: brute force
  coherence and non stationary noise analysis Tech. rep. LSC-Virgo Meeting,
  Pasadena, CA {LIGO} Document Number LIGO-G1500230

\bibitem{vajente2008analysis}
Vajente G 2008 {\em Analysis of sensitivity and noise sources for the Virgo
  gravitational wave interferometer\/} Ph.D. thesis Scuola Normale Superiore di
  Pisa, 2008. 11, 12, 13, 14, 21, 35, 44

\bibitem{hastie_09_elements-of.statistical-learning}
Hastie T, Tibshirani R and Friedman J 2009 {\em The elements of statistical
  learning: data mining, inference and prediction\/} 2nd ed (Springer)
  \urlprefix\url{http://www-stat.stanford.edu/~tibs/ElemStatLearn/}

\bibitem{ridge}
Hoerl A~E and Kennard R~W 1970 {\em Technometrics\/} {\bf 12} 55--67
  (\textit{Preprint}
  \eprint{https://www.tandfonline.com/doi/pdf/10.1080/00401706.1970.10488634})
  \urlprefix\url{https://www.tandfonline.com/doi/abs/10.1080/00401706.1970.10488634}

\bibitem{tibshirani1996regression}
Tibshirani R 1996 {\em Journal of the Royal Statistical Society. Series B
  (Methodological)\/} {\bf 58} 267--288

\bibitem{gwpy}
Macleod D, Areeda J, Urban A~L and Coughlin S 2018 gwpy/gwpy: 0.12.1
  \urlprefix\url{https://doi.org/10.5281/zenodo.1421570}

\bibitem{scikit-learn}
Pedregosa F, Varoquaux G, Gramfort A, Michel V, Thirion B, Grisel O, Blondel M,
  Prettenhofer P, Weiss R, Dubourg V, Vanderplas J, Passos A, Cournapeau D,
  Brucher M, Perrot M and Duchesnay E 2011 {\em Journal of Machine Learning
  Research\/} {\bf 12} 2825--2830

\bibitem{Luomala2015EffectsOT}
Luomala J and Hakala I 2015 {\em 2015 Federated Conference on Computer Science
  and Information Systems (FedCSIS)\/}  1247--1255

\bibitem{o1sensitivity}
Martynov D~V, Hall E~D {\em et~al.\/} 2016 {\em Phys. Rev. D\/} {\bf 93}(11)
  112004 \urlprefix\url{https://link.aps.org/doi/10.1103/PhysRevD.93.112004}

\bibitem{SEI_RTD_Investigation}
Derosa R 2016 Back to the usual excess noise
  \url{https://alog.ligo-la.caltech.edu/aLOG/index.php?callRep=24541}

\bibitem{SEI_RTD_Investigation_Detchar}
Lundgren A 2016 Could {PEM} have tracked down the {SEI RTD} noise?
  \url{https://alog.ligo-la.caltech.edu/aLOG/index.php?callRep=24625}

\bibitem{etmyalogjosh}
Smith J 2017 More on llo range variations - 60-70hz bump driven by ey ahu/temp
  \url{https://alog.ligo-la.caltech.edu/aLOG/index.php?callRep=31801}

\bibitem{etmyalogrobert}
Schofield R 2016 Tabletop measurements support p-cal camera mirror theory for
  the llo 65 hz peak, thermal expansion stick-slips may drive the peak, and a
  damping demonstration
  \url{https://alog.ligo-la.caltech.edu/aLOG/index.php?callRep=30422}

\bibitem{gwsumm}
Macleod D 2018 {Gravitational-wave network summary pages}
  \urlprefix\url{https://www.gw-openscience.org/detector_status/}

\bibitem{james2013introduction}
James G, Witten D, Hastie T and Tibshirani R 2013 {\em An introduction to
  statistical learning\/} vol 112 (Springer)

\bibitem{smith2006linear}
Smith J, Ajith P, Grote H, Hewitson M, Hild S, L{\"u}ck H, Strain K~A, Willke
  B, Hough J and Danzmann K 2006 {\em Classical and Quantum Gravity\/} {\bf 23}
  527

\bibitem{ranathesis}
Adhikari R 2004 {\em Sensitivity and Noise Analysis of 4 km Laser
  Interferometric Gravitational Wave Antennae\/} Ph.D. thesis Massachusetts
  Institute of Technology

\bibitem{székely2007}
Székely G~J, Rizzo M~L and Bakirov N~K 2007 {\em Ann. Statist.\/} {\bf 35}
  2769--2794 \urlprefix\url{https://doi.org/10.1214/009053607000000505}

\bibitem{richards2017distance}
Richards D~S~P 2017 {\em Notices of the American Mathematical Society\/} {\bf
  64} \urlprefix\url{https://arxiv.org/abs/1709.06400}

\bibitem{channelnaming}
Sigg D 1999 Channel naming convention Tech. rep. {LIGO} Document Number
  LIGO-T990033 \urlprefix\url{https://dcc.ligo.org/LIGO-T990033/public}

\end{thebibliography}

\end{document}